\documentclass[twocolumn]{aastex631}
\pdfoutput=1

\let\tablenum\relax
\usepackage[separate-uncertainty=true, range-phrase=--, range-units=single,
angle-symbol-over-decimal, retain-explicit-plus]{siunitx}
\usepackage{amsmath}
\usepackage{graphicx}
\usepackage{chemformula}
\usepackage{acro}
\acsetup{patch/longtable=false}

\received{}

\shorttitle{}
\shortauthors{}

\DeclareSIUnit{\mag}{mag}
\DeclareSIUnit{\pixel}{pixel}
\DeclareSIUnit{\parsec}{pc}
\DeclareSIUnit{\arcsec}{arcsec}
\DeclareSIUnit{\arcmin}{arcmin}
\DeclareSIUnit{\solarlum}{\mbox{$L_\odot$}}
\DeclareSIUnit{\solarmass}{\mbox{$M_\odot$}}
\DeclareSIUnit{\solarmetal}{\mbox{$Z_\odot$}}
\DeclareSIUnit{\year}{yr}
\DeclareSIUnit{\deg}{deg}
\DeclareSIUnit{\erg}{erg}

\newcommand{\ha}{H$\alpha$}
\newcommand{\hb}{H$\beta$}
\newcommand{\oiii}{[\ion{O}{3}]}

\newcommand{\oii}{[\ion{O}{2}]}

\newcommand{\hii}{\ion{H}{2}}
\newcommand{\hi}{\ion{H}{1}}

\DeclareAcroEnding{possessive}{'s}{'s}

\NewAcroCommand\acg{m}{\acropossessive\UseAcroTemplate{first}{#1}}

\DeclareAcronym{ew}{short=EW, long=equivalent width}
\DeclareAcronym{2mass}{short=2MASS, long=Two Micron All Sky Survey}
\DeclareAcronym{cigale}{short=\textsc{cigale}, long=Code Investigating GALaxy Emission}
\DeclareAcronym{sfr}{short=SFR, long=star formation rate}
\DeclareAcronym{psf}{short=PSF, long=point spread function}
\DeclareAcronym{agn}{short=AGN, long=active galactic nucleus, long-plural-form=active galactic nuclei}
\DeclareAcronym{XUV}{short=XUV, long=extended-UV}
\DeclareAcronym{sfe}{short=SFE, long=star formation efficiency}
\DeclareAcronym{lsb}{short=LSB, long=low surface brightness}
\DeclareAcronym{ism}{short=ISM, long=interstellar medium}
\DeclareAcronym{sdss}{short=SDSS, long=Sloan Digital Sky Survey}
\DeclareAcronym{cwru}{short=CWRU, long=Case Western Reserve University}
\DeclareAcronym{dig}{short=DIG, long=diffuse ionized gas}
\DeclareAcronym{ifs}{short=IFS, long=integral field spectroscopy}
\DeclareAcronym{imf}{short=IMF, long=initial mass function}
\DeclareAcronym{sfh}{short=SFH, long=star formation history, long-plural-form=star formation histories}
\DeclareAcronym{igm}{short=IGM, long=intergalactic medium}
\DeclareAcronym{hst}{short=\emph{HST}, long=\emph{Hubble Space Telescope}}
\DeclareAcronym{sed}{short=SED, long=spectral energy distribution}
\DeclareAcronym{galex}{short=\emph{GALEX}, long=\emph{Galaxy Evolution Explorer}}

\newcolumntype{P}[1]{>{\centering\arraybackslash}p{#1}}

\begin{document}

\title{A Dynamic Galaxy: Stellar Age Patterns Across the Disk of M101}
\author{Ray Garner, III}
\affiliation{Department of Astronomy, Case Western Reserve University, 10900 Euclid Avenue, Cleveland, OH 44106, USA}
\affiliation{Department of Physics and Astronomy, Texas A\&M University, 578 University Drive, College Station, TX, 77843, USA}

\author{J. Christopher Mihos}
\affiliation{Department of Astronomy, Case Western Reserve University, 10900 Euclid Avenue, Cleveland, OH 44106, USA}

\author{Paul Harding}
\affiliation{Department of Astronomy, Case Western Reserve University, 10900 Euclid Avenue, Cleveland, OH 44106, USA}

\author{Charles R. Garner, Jr.}
\affiliation{Rockdale Magnet School for Science and Technology, 930 Rowland Road, Conyers, GA, 30012, USA}

\begin{abstract}

Using deep, narrowband imaging of the nearby spiral galaxy M101, we present stellar age information across the full extent of the disk of M101. Our narrowband filters measure age-sensitive absorption features such as the Balmer lines and the slope of the continuum between the Balmer break and \SI{4000}{\angstrom} break. We interpret these features in the context of inside-out galaxy formation theories and dynamical models of spiral structure. We confirm the galaxy's radial age gradient, with the mean stellar age decreasing with radius. In the relatively undisturbed main disk, we find that stellar ages get progressively older with distance across a spiral arm, consistent with the large-scale shock scenario in a quasi-steady spiral wave pattern. Unexpectedly, we find the same pattern across spiral arms in the outer disk as well, beyond the corotation radius of the main spiral pattern. We suggest that M101 has a dynamic, or transient, spiral pattern with multiple pattern speeds joined together via mode coupling to form coherent spiral structure. This scenario connects together the radial age gradient inherent to inside-out galaxy formation with the across-arm age gradients predicted by dynamic spiral arm theories across the full radial extent of the galaxy.

\end{abstract}

\section{Introduction}


Spiral galaxies are believed to have formed ``inside-out,'' that is the inner parts of galactic disks form first followed by the formation of their outer regions \citep[e.g.][]{white1991,mo1998}. One notable consequence of inside-out formation is a radial variation in the star formation history, and thus stellar ages, of a spiral galaxy. These have been primarily observed as a color gradient in the disk, in that inner regions appear redder (older) than the bluer (younger) outer regions \citep[e.g.][]{dejong1996_color,bell2000,macarthur2004}. Complicating this picture are non-axisymmetric structures in the disk, such as spiral arms, that both radially scatter stars \citep[e.g.][]{sellwood2002,roskar2008_migration,roskar2008} and imprint azimuthal variations in the stellar age distribution \citep[e.g.][]{dixon1971,dobbs2010,chandar2017}.  

Understanding the nature and creation mechanism of spiral structure in galaxies is still a fundamental problem in astronomy. We lack a complete, and widely accepted, theory for the origin of spiral patterns. Many spiral galaxies have structure driven by non-axisymmetric effects, whether that be caused by external interactions with a companion \citep{kormendy1979,donghia2016,pettitt2016} or by internal responses to a centrally rotating bar \citep{contopoulos1980,athanassoula1992}. There is also the possibility that some spiral structure is self-excited. Broadly speaking there are two possible mechanisms for self-excitation. First, there is the ``density wave theory'' that argues spiral structure is the result of a quasi-steady, global mode in the stellar disk \citep{lin1964,bertin1989}. Second, there is the suggestion that spiral structure is caused by transient and recurrent instabilities resulting in ``dynamic'' spiral arms that appear and reappear in cycles \citep{toomre1964,sellwood1984,elmegreen1986,sellwood1991}. We will refer the interested reader to any number of reviews on spiral mechanisms for more detailed information \citep[e.g.][]{dobbs2014_dawes,shu2016,sellwood2022}. 

\begin{figure}
\plotone{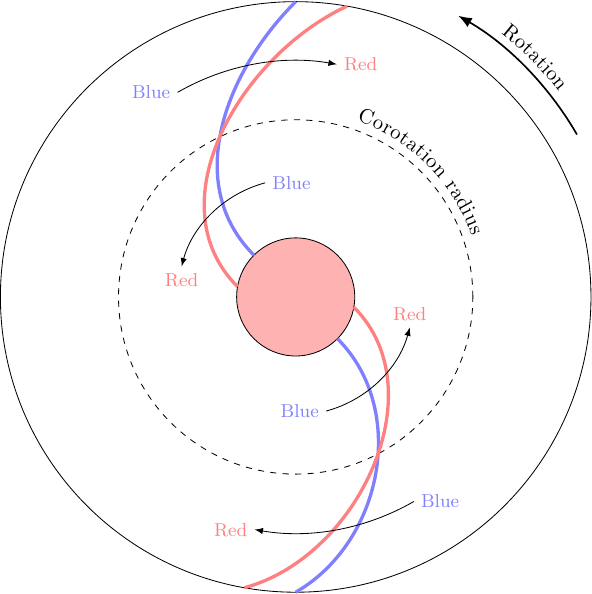}
\caption{Schematic of the shock scenario posited by \citet{roberts1969}. Stellar age gradients across the spiral arms are indicated by arrows that go from blue to red. The azimuthal age gradients are produced by stars born in the spiral shock, where the shocked interstellar medium forms a dust lane, and later drift away as they age. The direction of the gradients flips at the corotation radius. Inside corotation, the gas and stars overtake the spiral pattern, and outside corotation the spiral pattern overtakes the gas and stars.}
\label{spiral_shock} 
\end{figure}

Clearly there are multiple spiral pattern generation mechanisms and we must turn to observations and simulations of spiral galaxies to inform us which is the primary mechanism. However, as highlighted by \citet{sellwood2011}, it is very difficult to find observational tests to distinguish between these theoretical models. As an example, in the standard density wave theory, spiral arms are density waves moving with a constant pattern speed. Inside corotation (the radius at which the angular speed of stars and gas equals the pattern speed), material will rotate faster than the spiral pattern. When gas enters the spiral pattern, it may experience a shock and collapse to form stars (\citealt{roberts1969,shu1972}; see \citealt{mckee2007} and references therein for an extensive discussion of the theory). As these stars age, they overtake the more slowly rotating spiral pattern, drifting away from their birth sites and creating an azimuthal age gradient across the spiral arm (Figure~\ref{spiral_shock}; also Figure~1 in \citealt{martinezgarcia2009}). We would then expect to find young star clusters near the spiral arm along the leading edge and older stars further along, approaching the trailing side of the next spiral arm. Outside of corotation, the spiral pattern moves faster than the gas and the opposite sequence occurs \citep{dixon1971}. 

Numerous simulations have been performed in recent years to test this theory. For example, \citet{dobbs2010} found that in galaxies with a constant pattern speed, a clear age sequence across spiral arms from young to old stars is expected. This pattern was absent in dynamic and tidally-disturbed galaxies. Subsequent simulations with more advanced hydrodynamical modeling have confirmed the case for the latter pair of spiral arm mechanisms \citep{wada2011,grand2012,donghia2013,baba2015,baba2017,dobbs2017,pettitt2017}. Observationally, evidence of the large-scale shock scenario is hard to come by except in a few cases. \citet{martinezgarcia2009} studied color gradients across the spiral arms of thirteen spiral galaxies, ten of which had the expected color gradient. They have since found color gradients for a few more galaxies \citep{martinezgarcia2011,martinezgarcia2015}. Similarly, \citet{sanchezgil2011} produced age maps of six nearby galaxies using the \ha/FUV flux ratio, only two of which (M74 and M100) presented an age gradient across spiral arms. Investigating resolved stellar clusters in three galaxies, \citet{abdeen2022} found evidence of the expected age gradients in galaxies where others have not. Simply put, with the exception of a handful of specific cases, there is no clear systematic trend in the observations.

Some of the observational ambiguity likely comes from the wide use of broadband colors to discern age gradients. The well-known age-metallicity degeneracy \citep{worthey1994_age}, which combined with the reddening effects of dust, blurs the conclusions made with color gradients. Spectroscopy gives the ability to break this degeneracy by resolving multiple age- and metallicity-sensitive absorption features. However spectroscopy becomes prohibitively expensive to measure the \ac{lsb} outskirts of disk galaxies. For nearby galaxies with large angular sizes---where we can resolve physically small scales---most integral field units are also not large enough to cover the whole galaxy without resource-intensive mosaicing techniques. 

Another alternative to broadband photometry is narrowband photometry, particularly observations targeting age-diagnostic absorption features like the Balmer lines, the \SI{4000}{\angstrom} break, or metallic lines such as Mg~b. Pioneering narrowband work has been performed for spiral galaxies \citep{beauchamp1997,molla1999,ryder2005} and modern narrowband photometry can measure absorption line strengths that are compatible with spectroscopy \citep{stothert2018,angthopo2020,renard2022}. Absorption line \acp{ew} are largely insensitive to dust effects \citep{macarthur2005} and, combined with stellar population synthesis modeling, have been used to investigate luminosity-weighted stellar ages in integrated stellar populations \citep[e.g.][]{fisher1995,ganda2007,sanchezblazquez2014a,sanchezblazquez2014b}

In an effort to study stellar ages and their connection to the nature of spiral patterns, we have used \acg{cwru} Burrell Schmidt 24/36-inch telescope and its accompanying narrowband filters to image the nearby spiral galaxy M101 (NGC~5457). While these images targeting \ha, \hb, \oiii$\lambda\lambda$4959,5007 and \oii$\lambda\lambda$3726,3729 have been used to study the emission line properties of M101 and its group environment \citep{watkins2017,garner2021,garner2022}, these images also reveal age-diagnostic absorption signatures in the stellar disk of M101 (Figure~\ref{abs_images}). Thus, by measuring equivalent widths through our narrowband filters, we are able to place constraints on the stellar ages throughout M101's disk. 

M101 was chosen for this survey because its nearby distance ($D = \SI{6.9}{\mega\parsec}$; see \citealt{matheson2012} and references therein) enables its properties to be studied in great detail at high spatial resolution. Indeed, \citet{lin2013} performed a pixel-based multiwavelength \ac{sed} fitting and broadly found that M101 supports the inside-out disk growth scenario with detections of radial stellar age and metallicity gradients. The disk of M101 also has a dynamic nature: it has strong morphological asymmetries \citep[e.g.][]{beale1969}, complex \hi\ kinematics \citep[e.g.][]{waller1997,mihos2012,xu2021}, a Type I \acl{XUV} disk (\citealt{thilker2007}), and faint tidal features to the northeast and east \citep{mihos2013,mihos2018}. These are all likely signatures of an interaction M101 had with its most massive satellite NGC~5474 about \SI{300}{\mega\year} ago \citep{linden2022}. Thus M101 provides an interesting, albeit often overlooked, testing ground for the varying roles of self-excited spiral patterns and tidally-induced spiral patterns. 

\section{Narrowband Observations}

The narrowband imaging data for this project was taken over four observing seasons using \ac{cwru} Astronomy's 24/36-inch Burrell Schmidt telescope located at Kitt Peak Observatory. Full details of our narrowband imaging techniques are given in \citet{watkins2017} and \citet{garner2022}. Quantitative information about the narrowband filters and final imaging stacks for the dataset is given in Table~2 of \citet{garner2022}. Briefly, we summarize our observations here. 

Our dataset consists of a set of narrow on-band filters ($\Delta\lambda \approx$ \SIrange{80}{100}{\angstrom}) centered on the redshifted emission lines \ha$\lambda$6563, \hb$\lambda$4861, \oiii$\lambda\lambda$4959,5007, and \oii$\lambda\lambda$3727,3729. To measure the adjacent stellar continuum for each line, we also observed M101 in narrow off-band filters shifted in wavelength by $\approx$\SIrange{100}{150}{\angstrom} from each on-band filter. The Burrell Schmidt images a $\ang{1.65}\times\ang{1.65}$ field of view onto a $4096 \times 4096$ back-illuminated CCD, yielding a pixel scale of \SI{1.45}{\arcsec\per\pixel}. In each filter, we took \numrange{50}{70} \SI{1200}{\second} images of the galaxy, randomly dithering the telescope by \SIrange{10}{30}{\arcmin} between exposures. Both twilight flats and night sky flats were taken throughout each observing run, as were images of spectrophotometric standards and bright stars to assist with photometric calibration and scattered light modeling, respectively. 

All observed images are treated to the data reduction procedures described by \citet{watkins2017} and \citet{garner2022}. Namely, the steps are as follows: (1) subtractions of bias frame; (2) flat-field correction; (3) removal of scattered light from bright stars following the technique of \citet{slater2009}; (4) registration of stacked images; (5) flux calibration of the stacked images. We flux calibrated the final image stacks using a variety of techniques: (1) deriving a photometric solution from observations of \citet{massey1988} spectrophotometric standard stars, (2) measuring zeropoints from $ugr$ magnitudes of the $\sim$150 stars in the M101 field from the \ac{sdss}, and (3), where possible, synthesizing narrowband magnitudes using \ac{sdss} spectroscopy of $\sim$100 point sources in the M101 field. These different techniques yielded flux zeropoints that agreed to within $\pm$\SI{5}{\percent}, which we take to be the absolute flux uncertainties in the data. 

\begin{figure*}
\gridline{\fig{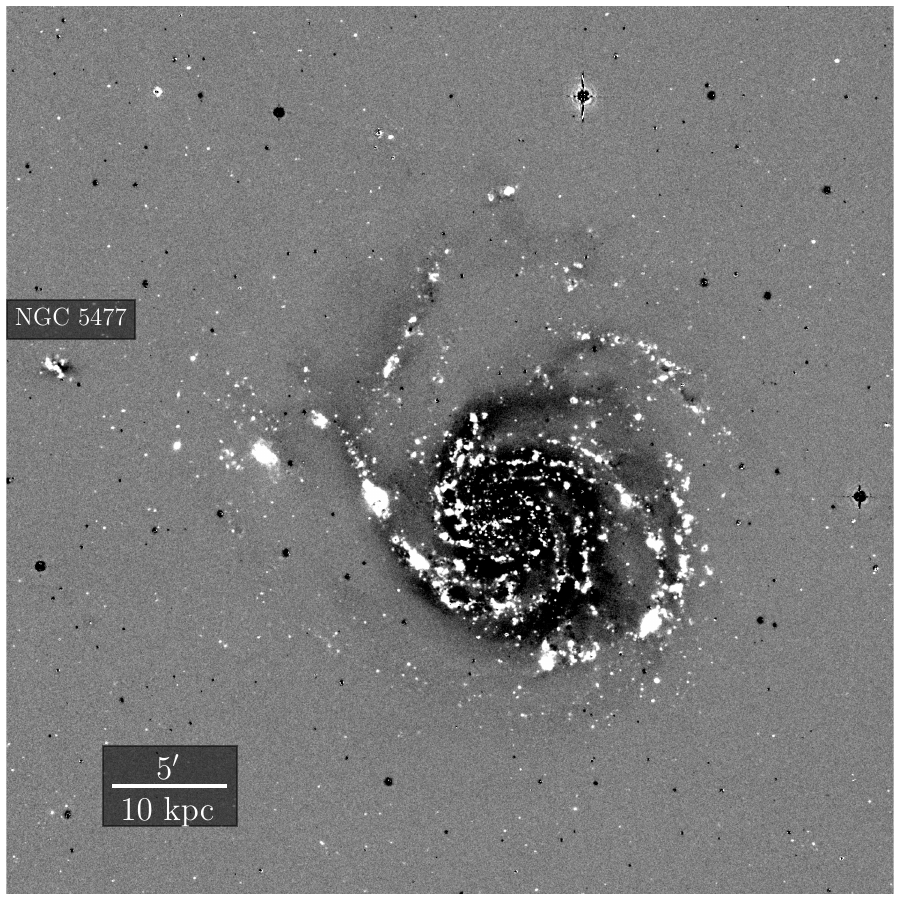}{0.5\textwidth}{(a)}
\fig{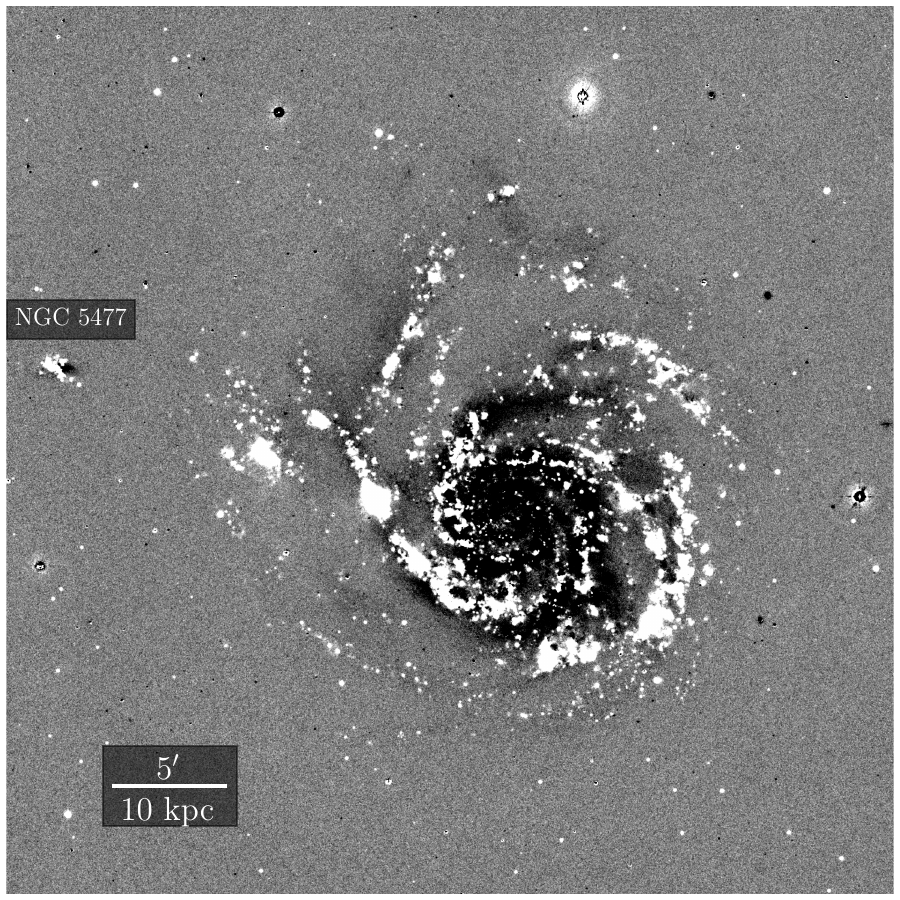}{0.5\textwidth}{(b)}}
\caption{(a) The continuum-subtracted \hb\ image. (b) The continuum-subtracted \oii\ image. Images are scaled in such a way that emission-line objects are in white and absorption features are in black. Note the strong absorption features in the disk of M101. The satellite galaxy NGC~5477 is marked to the left and a scale bar is provided. The image measures $\ang{;40;} \times \ang{;40;}$. North is up and east is to the left.}
\label{abs_images}
\end{figure*}

Previously we used these narrowband images of M101 and its group environment to investigate emission-line sources, whether those be in the intragroup environment \citep{garner2021} or within the disk of M101 and its satellites \citep{watkins2017,garner2022}. However the level of precision and accuracy of our imaging and reduction techniques reveals absorption signatures in the images, particularly in the \hb\ and \oii\ continuum-subtracted images. We focus on these two images out of our narrowband dataset since our \oiii\ filters do not sample any strong stellar absorption features and while trends in \ha\ absorption are similar to the \hb\ absorption, \ha\ absorption is less sensitive to age than \hb\ \citep[e.g.][]{gonzalezdelgado2005}. Figure~\ref{abs_images} shows the \hb\ and \oii\ continuum-subtracted images, and within the disk of M101 darker regions are present. We stress that this is not incorrect off-band over-subtraction in our data reduction techniques, but rather the presence of absorption signatures caused by the underlying older stellar population. 

In order to focus only on features in the stellar continuum, we need to mask both foreground stars and strong emission line \hii\ regions in M101. We follow the techniques described by \citet{garner2022} to remove foreground Milky Way stars from our images. We masked those stars in the Tycho-2 Catalog \citep{hog2000} brighter than $B_T = 12.5$. All of these stars lie projected outside of the disk of M101. For fainter stars, we used the \acl{2mass} All-Sky Catalog of Point Sources \citep{skrutskie2006}. We aggressively mask these stars using \ang{;;8.7} apertures (roughly three times the FWHM of the PSF of the coadded image stacks). To mask \hii\ regions, we used the segmentation map produced by \citet{garner2022} which used \texttt{astropy}'s \texttt{PhotUtil} package and its \texttt{segmentation} module \citep{bradley2023} to identify bright \hii\ regions using the continuum-subtracted \ha\ image. Finally, we also mask the nearby satellite galaxy NGC~5477, visible in our images projected \ang{;22;} or \SI{44}{\kilo\parsec} to the east.

We note that although we have masked the brightest portions of the strongly emitting \hii\ regions, there is still scattered emission and \ac{dig} in our images. Briefly, scattered emission refers to light from bright \hii\ regions scattered towards the observer by interstellar dust particles in M101's disk regardless of the ionization state of the gas (e.g.\ \citealt{brandt2012} and references therein), while the \ac{dig} refers to the warm, low density, ionized component of the \acl{ism} (\acs{ism}; see \citealt{haffner2009} and references therein). The \ac{dig} has a varying presence throughout a galaxy, and so we correct for the \ac{dig} using the approach of \citet{valeasari2019}. They separate the \ac{dig} from normal star-forming regions on the basis of the measured \ha\ \ac{ew}. Thus the required correction as prescribed by Equation 2 of \citet{valeasari2019} is $\sim$\SI{5}{\percent} in all of our narrowband observations, which we apply to our data. There is no similar correction for scattered light, but this will be strongest near \hii\ regions which we have aggressively masked. 

\begin{figure}
\epsscale{1.2}
\plotone{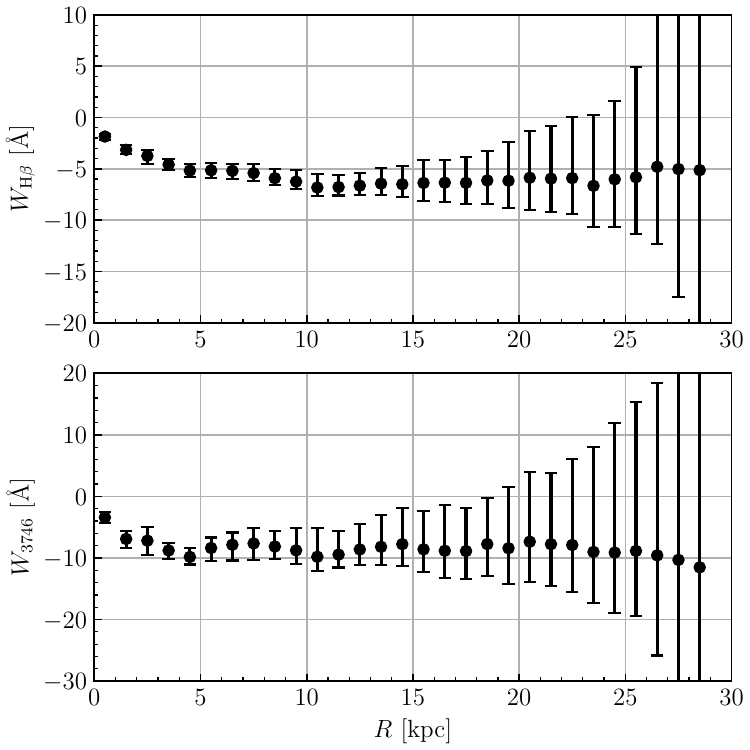}
\caption{The radial profiles of the median stellar \acp{ew} measured through our \hb\ and \oii\ filters, $W_{\mathrm{H}\beta}$ and $W_{3746}$, respectively. Radial bins have widths of \SI{1}{\kilo\parsec} (\ang{;;30}). The error bars indicate the quartile scatter of the \acp{ew} measured in each radial bin. }
\label{radial_prof}
\end{figure}

As an example of the extent and depth of our data, Figure~\ref{radial_prof} shows the binned radial profiles of the median stellar \ac{ew} (after masking \hii\ regions) measured through our \hb\ and \oii\ filters, $W_{\mathrm{H}\beta}$ and $W_{3746}$, respectively, out to \SI{30}{\kilo\parsec} (\ang{;15;}).\footnote{We note that $W_{3746}$ is named for the central wavelength of the on-band filter, not for the equivalent width of any individual absorption line.} Importantly, the error bars are dominated by physical scatter in data rather than measurement error. The \hb\ \ac{ew}, $W_{\mathrm{H}\beta}$, shows a negative gradient out to $R \sim \SI{10}{\kilo\parsec}$, beyond which it flattens and the \oii\ \ac{ew}, $W_{3746}$, shows a negative gradient out to $R \sim \SI{5}{\kilo\parsec}$, similarly flattening beyond this radius. Interpretation of these trends is simplest for the $W_{\mathrm{H}\beta}$, with the presence of A-type stars increasing with radius, subsequently deepening the \hb\ absorption line, although the scatter at large radius might suggest the presence of younger populations. However, the interpretation of the \ac{ew} measured through the \oii\ filters is less straightforward and motivates more detailed spectral modeling. Additionally, the large variance in \acp{ew} in the outer disk motivates treating the main and outer disks separately, where in the main disk we can explore any trends with respect to morphological environments. 

\section{Defining Environments Within M101's Disk}\label{sec:masks}

Our goal is to construct morphological, environmental masks to allow us to study the stellar ages of different regions in M101's disk. In order to outline stellar structures, we utilize the broadband $B$ image of M101 \citep{mihos2013}. In the following, we define several basic environments that are included in the masks: inner disk, spiral arms and interarm regions, and the main and outer disks. 

\subsection{The Inner Disk}

In traditional photometric decompositions of galaxies, the exponential disk is usually distinguished from a central bulge component. Additionally, some galaxies have unresolved or marginally resolved stellar structures that are centrally concentrated. Some of these structures might be an unresolved nuclear bar, a nuclear ring, or a nuclear disk. Finally, at small radii, spiral arms and their interarm regions become increasingly hard to define in an objective manner. 

In the specific case of M101, its morphological classification is that of a spiral galaxy without a large classical bulge (S\underline{A}B(rs)c; \citealt{buta2015}). It is known to have a mildly star-forming pseudobulge \citep{fisher2009,fisher2010,kormendy2010} with an effective radius of \SI[multi-part-units=single]{372 \pm 436}{\parsec} ($\ang{;;11.1} \pm \ang{;;13.0}$; \citealt{fisher2010}). While other studies have resolved and studied the stellar age of the pseudobulge and surrounding region \citep{lin2013}, since our focus is on the spiral arms and interarm regions, we visually define an inner disk region of \ang{;;120} (\SI{4}{\kilo\parsec}). This encompasses the pseudobulge and the region where the spiral arms become so tightly wrapped they are hard to define. We note that this is much larger than what studies of morphological features have considered to be the central region, often $<\ang{;;20}$ \citep{lin2013,querejeta2021}.

\subsection{Spiral Arms and Interarm Regions}

In general, logarithmic spirals are good approximations to the shape of galactic spiral arms \citep{seigar1998}. Ubiquitous throughout nature, logarithmic spirals are easy to mathematically define (see Section 2 of \citealt{davis2017}), but defining in a robust sense where spiral arms are in a galaxy is more difficult. Over the years, numerous methods have been presented in the literature to define spiral arms and estimate their pitch angles (see \citealt{hewitt2020} and references therein). These methods, while useful for large surveys due to their (semi-)automation, each come with their own issues. For instance, some methods may not always trace real three- or four-armed spiral patterns in galaxies, nor do they handle asymmetric arms well \citep{elmegreen1992}. These methods also generally only model one global pitch angle for a galaxy, rather than accurately define a more complicated spiral pattern.

Since our focus is on M101 and its strongly asymmetric spiral pattern, we forego these automated methods and instead adopt a different approach based on the visual inspection of the spiral arms. This allows for multiple pitch angles at different radii in M101, and takes into account that those pitch angles will be asymmetrical with respect to the galaxy center. Thus we follow a semi-automatic procedure developed by the S$^4$G \citep{sheth2010} and PHANGS \citep{lee2022} teams in \citet{herreraendoqui2015} and \citet{querejeta2021}, respectively. We briefly summarize those procedures and our slight modifications here. 

First we create an unsharp mask using the $B$-band image from \citet{mihos2013}. This is done by convolving the image with a Gaussian kernel and then dividing the original image with the smooth convolved image \citep{malin1977}. The width of the Gaussian kernel was chosen to be 30 pixels (\ang{;;43.5}). This unsharp mask highlights the spiral features in the disk of M101. Points along each spiral arm are marked in \texttt{SAOImage ds9} \citep{ds9} and their coordinates are then transformed into logarithmic polar coordinates, $(\ln(r), \theta)$, where true logarithmic spirals appear as straight lines. 

\begin{figure*}
\plotone{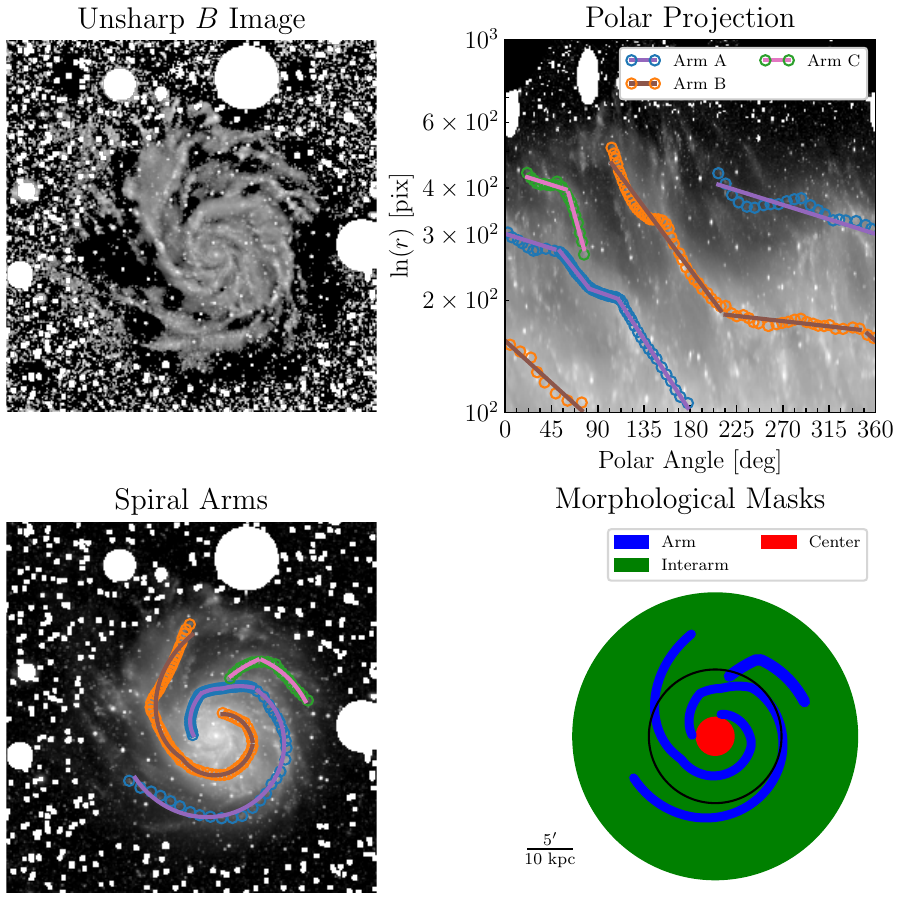}
\caption{An illustration of the steps of our spiral arm detection method. Top left: the unsharp, masked $B$-band image. Points were marked on this image to define the spiral arms. Top right: the unsharp, masked $B$-band image in log-polar coordinates. The different colored points and lines show the spiral arm segments that were fitted with different pitch angles. Bottom left: the masked $B$-band image with the spiral arm segments overlaid on top. Bottom right: an illustrative schematic of the morphological masks used. The black circle marks the boundary between the main and outer disk at \ang{;;430} (\SI{14.4}{\kilo\parsec}). See text.}
\label{morph_masks}
\end{figure*}

It is here that we deviate slightly from previous works. To allow for the pitch angle to vary as a function of radius, we start by making a preliminary visual estimate of the arm segments in log-polar coordinates. We then pass these starting values into our code to refine the number of break points in the data. The positions of those break points are estimated iteratively using the break-fitting method outlined in \citet{muggeo2003} and implemented with the \texttt{Python} package \texttt{piecewise-regression} \citep{piecewise}. The algorithm randomly generates a series of break points and fits a model of a line with a term that incorporates a change in gradient between some number of segments of a piecewise function. Using ordinary linear regression and bootstrap resampling \citep{wood2001}, this process is iterated until the position of the break converges. The results of this semi-automatic process are then projected back into the plane of the sky. 

We then need to define a width of each of our spiral arms. Since we want to differentiate between the star-forming spiral arms and the less active interarm regions, we use the continuum-subtracted \ha\ image to iteratively dilate the spiral arm until the measured \ha\ flux within the arm reaches an empirical threshold following the procedure outlined in \citet{querejeta2021}. This process results in widths of \SI{2}{\kilo\parsec} to \SI{2.5}{\kilo\parsec}, encompassing most of the \ha\ emission that one would associate with an arm by eye.

Finally, we defined the interarm regions. For simplicity, the interarm regions are simply any part of the galaxy not in the spiral arms or inner disk. This does include the extreme outer disk of M101 where there are no spiral arms but still low levels of star formation. Figure~\ref{morph_masks} shows the spiral arm and interarm masks, as well as an illustration of the process described above.

\subsection{Main and Outer Disks}

In addition to the morphological masks described above, we also make a distinction between the main and outer disks of M101. This is motivated by two reasons. First, the outer disk of M101 has structures that are quite irregular and likely tidally-induced in nature while the main disk is relatively ordered. We would then expect the outer disk to be more disorganized in terms of any stellar age gradients. Second, the high surface brightness main disk has higher signal-to-noise than the outer disk, and, unlike the outer disk, can be studied without the need for large-scale rebinning of the pixel data. While both disk regions will be analyzed in similar ways, the low surface brightness of the outer disk requires a different treatment to build up signal. 

Therefore we define a boundary of \ang{;;430} (\SI{14.4}{\kilo\parsec}), roughly three times the disk scale length \citep{mihos2013,watkins2017}. We define the main disk to be inside this boundary, including the inner disk and portions of the spiral arms and interarm regions. We define the outer disk to be outside this boundary to a radius of \ang{;;920} (\SI{30.8}{\kilo\parsec}). The maximum radius was chosen to represent where the sky noise starts to dominate in our images. This is also approximately the outermost radius at which \hii\ regions were detected in \citet{garner2022}. 

Finally, in order to maximize the signal-to-noise in our images, we create a binned version of our narrowband images. Using the masked images, for studying the main disk we bin the image into $9 \times 9$ pixel ($\ang{;;13} \times \ang{;;13}$ or $\SI{450}{\parsec} \times \SI{450}{\parsec}$) blocks, calculating the median intensity of unmasked pixels in each block. In order to quantify uncertainties for each of these blocks, we also calculate the uncertainty estimate of the median \citep{rider1960,williams2001},
\begin{equation}
	\sigma_{\text{median}} = \sigma_{\text{mean}}\sqrt{\frac{\pi}{2}} = \text{std}\sqrt{\frac{\pi}{2N}},
\end{equation}
where $\text{std}$ is the standard deviation, $N$ is the number of pixels in a block, and $\sigma_{\text{mean}}$ is the uncertainty in the mean. However, when studying the outer disk, we bin over much larger scales described later in Section~\ref{sec:outer_disk}. It is from these medianed blocks that we calculate fluxes and \acp{ew} in each of the environments described above. In order to further maximize the signal-to-noise, in the main disk region we reject low signal-to-noise pixels at a threshold that corresponds to a surface brightness of roughly $\mu_B = \SI{23.6}{\mag\per\square\arcsec}$.

\section{Inferring Ages with Absorption Lines}\label{sec:models}

\begin{figure*}
\plotone{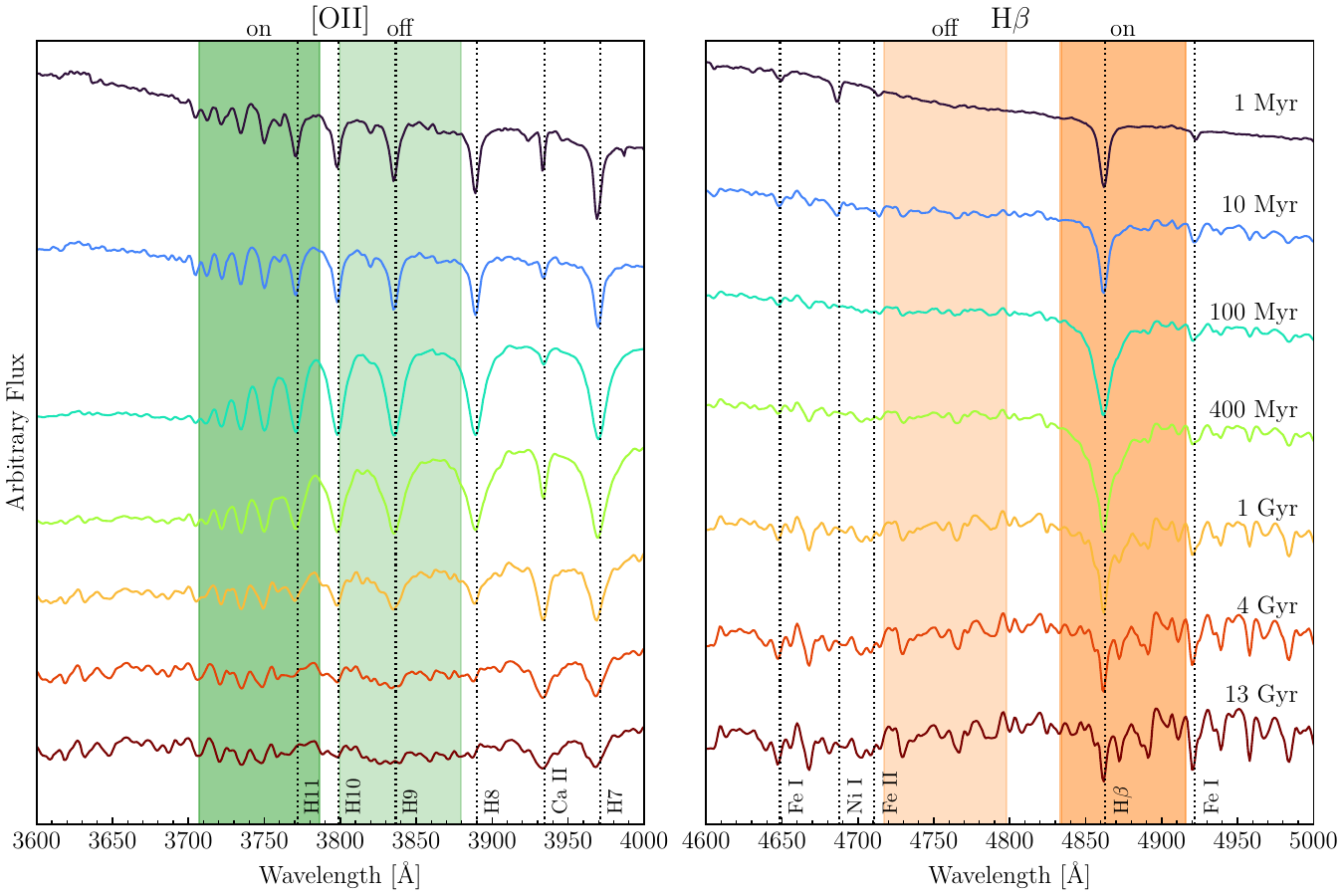}
\caption{The spectral energy distribution of an instantaneous burst of star formation at solar metallicity and with a \citet{chabrier2003} IMF, increasing in age from top to bottom: \SI{1}{\mega\year}, \SI{10}{\mega\year}, \SI{100}{\mega\year}, \SI{400}{\mega\year}, \SI{1}{\giga\year}, \SI{4}{\giga\year}, and \SI{13}{\giga\year}. Left: the \oii\ filter set. Right: the \hb\ filter set. In both panels, the on-band is darker than the off-band. \acp{sed} created with \ac{cigale} \citep{noll2009,boquien2019}.}
\label{simple_pop}
\end{figure*}

To connect our measured narrowband imaging to the underlying stellar populations, we utilize the SED fitting and modeling \acl{cigale} (\acs{cigale}; \citealt{noll2009,boquien2019}). To illustrate how our filters span various spectral features, Figure~\ref{simple_pop} shows the \ac{sed} of an instantaneous burst of star formation with solar metallicity that evolves to late times as generated with \ac{cigale}. Two broad spectral features are covered by our filters: individual Balmer lines in both the \hb\ and \oii\ filters, and in the \oii\ filter the slope of the blue continuum between the Balmer break at \SI{3646}{\angstrom} and the \SI{4000}{\angstrom} break, produced primarily by A/F and O/B stars, respectively. It is important to note that we cannot make a standard $D_{4000}$ measurement as both of our \oii\ filters are bluer than the wavelengths typically used for the $D_{4000}$ measurement \citep{bruzual1983,balogh1999}.

Our \hb\ filters are most sensitive to the strengths of their respective Balmer lines, while our \oii\ filters are sensitive to the strengths of the higher-order Balmer lines, and the slope of the continuum between the Balmer break and the \SI{4000}{\angstrom} break. Fortuitously, the \hb\ filters are positioned in such a way that they avoid measuring the nearby metallic lines, which reduces the chance of any secondary metallicity effects in the Balmer lines. On the other hand, the \oii\ filters act as a sort of ``pseudo-measurement'' of the Balmer break and the \SI{4000}{\angstrom} break, targeting no specific feature but sensitive to all of them. This motivates the need for stellar population modeling to understand how populations of different ages, metallicities, dust absorption, etc., behave through our filters. 

\subsection{Stellar Population Parameters}

In order to understand how the narrowband spectral features progress with age and what stellar population parameters they are sensitive to, we created a grid of model \acp{sfh} using \ac{cigale}. While \ac{cigale} is most commonly used to fit the observed multiwavelength \ac{sed} of galaxies, we instead use \ac{cigale} as a simple spectral synthesis code to model the observed \acp{ew} in each filter pair under a variety of \acp{sfh}. In this way, we can use our understanding of the models to inform our understanding of the dataset. In the following, we describe the parameters that we investigated. The input parameters and their values used are listed in Table \ref{cigale-params}. \ac{cigale} uses theoretical models to parameterize the flux emitted and absorbed by the stars, gas, and dust in model galaxies and produces a grid of model spectra that are then converted to \acp{sed}. \citet{boquien2019} describes the input models in complete detail, and we briefly review each chosen component and its contribution. 

To build a model galaxy \ac{sed}, we first need to define the properties of the underlying stellar population in terms of an assumed stellar population synthesis model, an \ac{imf}, and \ac{sfh}. We adopt here in all models the high-resolution version of the \citet{bruzual2003} stellar population synthesis models. We also adopt the \citet{chabrier2003} \ac{imf} with solar metallicity ($Z_\odot = 0.02$). While M101's average stellar metallicity is somewhat subsolar ($Z = \SI[multi-part-units=single]{0.5 \pm 0.3}{\solarmetal}$; \citealt{lin2013}), the relative insensitivity of the line strengths to metallicity in our filters make this a reasonable choice. However, we do investigate the effect of subsolar ($Z = 0.008$, $0.4Z_\odot$) and supersolar ($Z = 0.05$, $2.5Z_\odot$) metallicities on the generated models in the next section.

\begin{deluxetable*}{m{6cm} c P{9cm}}
\tablecaption{Input Parameters for SED Model Grid \label{cigale-params}}

\tablehead{\colhead{Parameter} & \colhead{Symbol} & \colhead{Range}}
\startdata
IMF: \citet{chabrier2003} & & \\
Metallicity & $Z$ & \numlist[list-final-separator = {, }]{0.008;0.02;0.05} \\
Redshift & $z$ & \num{0.0008} \\ \hline
\underline{SFH: \texttt{sfh2exp}} & & \\ 
Age of main population & $t$ & \SIlist[list-units = single, list-final-separator = {, }]{0.01; 0.02; 0.03; 0.04; 0.05; 0.06; 0.07; 0.08; 0.09; 0.1; 0.15; 0.2; 0.4; 0.6; 0.8; 1; 1.2; 1.4; 1.6; 1.8; 2; 2.3; 2.5; 2.7; 2.9; 3.1; 3.5; 3.7; 3.9; 4.1; 4.3; 4.5; 4.7; 4.9; 5.1; 5.3; 5.5; 5.7; 5.9; 6.1; 6.3; 6.5; 6.7; 6.9; 7.1; 7.3; 7.6; 7.8; 8; 8.2; 8.4; 8.6; 8.8; 9; 9.2; 9.4; 9.6; 9.8; 10}{\giga\year}\\
$e$-folding timescale of main population & $\tau$ & \SIlist[list-units = single, list-final-separator = {, }]{0.1;0.3;0.5;1;2;3;4;5;10;15;20;30}{\giga\year} \\
Age of a recent burst & $t_b$ & \SIlist[list-units = single, list-final-separator = {, }]{10;50;100;500}{\mega\year}\\
$e$-folding timescale of recent burst & $\tau_b$ & \SIlist[list-units = single, list-final-separator = {, }]{50}{\mega\year}\\
Burst fraction of total mass & $f_b$ & \numlist[list-final-separator = {, }]{0.0;0.01;0.1;0.5;0.7;0.9}\\ \hline
\underline{Dust Attenuation: \texttt{dustatt\_modified\_CF00}} & & \\
$V$-band attenuation in the ISM & $A_{V,\text{ISM}}$ & \numlist[list-final-separator = {, }]{0.0;0.1;0.3;0.5} \\
Fraction of total effective optical depth contributed by ISM & $\mu$ & \num{0.44} \\
Power law slope of dust attenuation in the ISM & $\delta_{\text{ISM}}$ & \num{-0.7} \\
Power law slope of dust attenuation in the birth clouds & $\delta_{\text{BC}}$ & \num{-1.3} 
\enddata
\end{deluxetable*}

\begin{figure}
\plotone{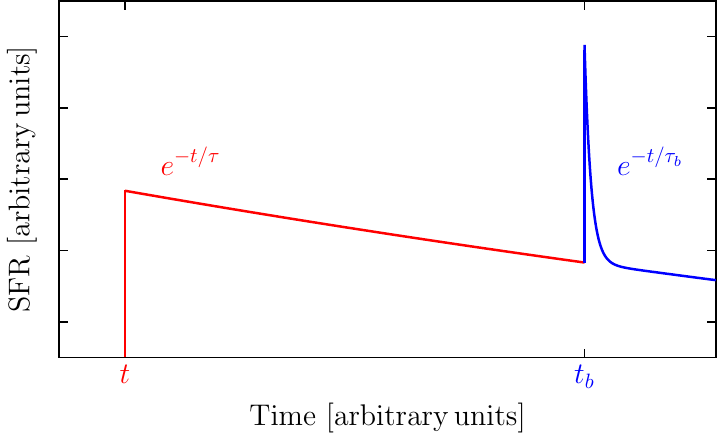}
\caption{Schematic of the modeled star-formation histories used. Red lines and text show the main population while blue lines and text show the recent burst population. The grids samples the labeled parameters (e.g., $t$, $t_b$, $\tau$, and $\tau_b$) over reasonable ranges. See text.}
\label{sfr_schematic}
\end{figure}

We model the \ac{sfh} as one or two decaying exponentials (Figure~\ref{sfr_schematic}). The first exponential models the long-term star formation that has formed the underlying stellar mass of the disk, while the optional second exponential models a more recent burst of star formation as might be expected due to spiral arm passages or tidal encounters. Both exponentials are parameterized by the $e$-folding times of the old and young populations ($\tau$ and $\tau_b$, respectively), the time since the beginning of each star formation model for the old and young populations ($t$ and $t_b$, respectively), and the fraction of stars formed in the second burst relative to the total mass of stars ever formed, $f_b$. The standard values we take for the $e$-folding times are $\tau = \SI{30}{\giga\year}$ (i.e., mimicking a slowly declining star formation rate) and $\tau_b = \SI{100}{\mega\year}$. For reference, under exponential \ac{sfh} models, typical $e$-folding timescales for spiral galaxies range from \SIrange{2}{30}{\giga\year} \citep{bolzonella2000}. 

Dust attenuation is handled following the \citet{charlot2000} model. This model makes the distinction between the heavily-extinguished stellar birth clouds and the less extinguished ambient \ac{ism}. We vary the $V$-band attenuation of the ambient \ac{ism} which also proportionally varies the $V$-band attenuation of the stellar birth clouds (see Equation~7 in \citealt{boquien2019}). For reference, \citet{lin2013} derived an integrated, global extinction for M101 of $A_{V} = 0.24$, although the central region is dustier, $A_{V} = 0.41$ (see also \citealt{boissier2004}). For our study, since we are masking the dense star-forming \hii\ regions before analysis, we expect typical extinction values for the regions we study to be significantly lower. 

\subsection{Diagnostic $W_{3746}$ vs.\ $W_{\mathrm{H}\beta}$ Plots}\label{sub:pop_explain}

\begin{figure*}
\plotone{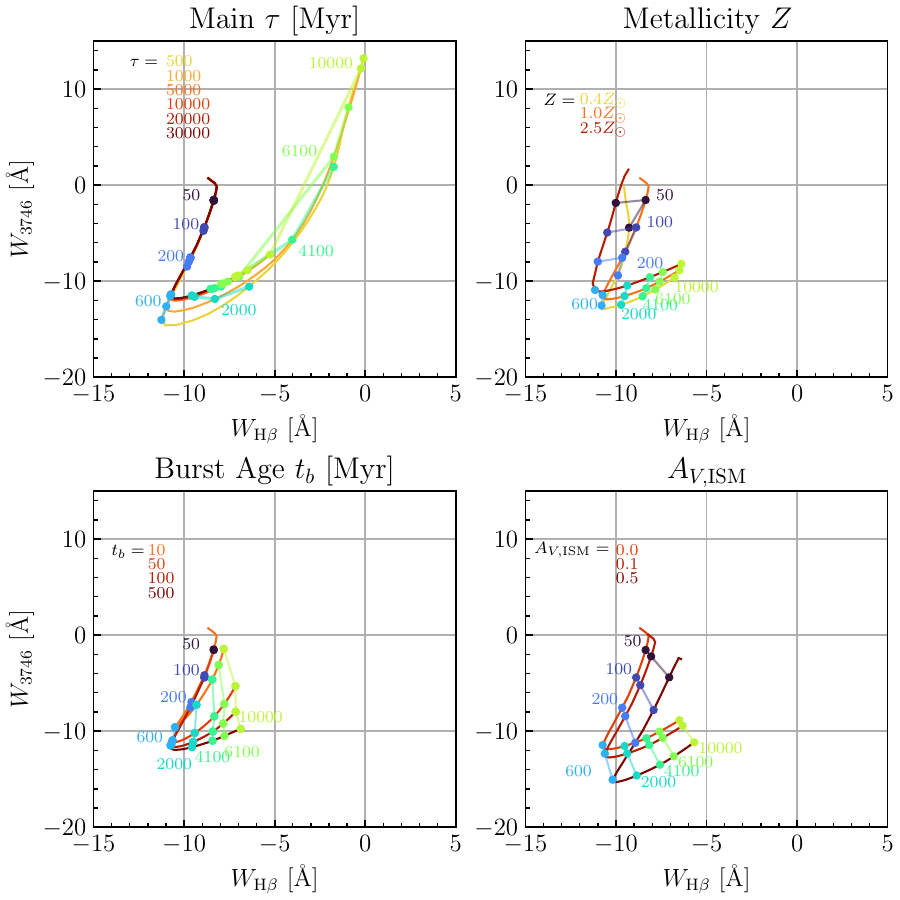}
\caption{Models for $W_{3746}$ vs.\ $W_{\mathrm{H}\beta}$ under different parameter assumptions using \ac{cigale}. Strong absorption is to the bottom left in each plot. All models assume a \citet{chabrier2003} \ac{imf}. Colored points along each track indicate the age of the population in \si{\mega\year}. Top left: varying the main population $\tau$, assuming solar metallicity and no recent burst. Top right: varying the metallicity, assuming $\tau = \SI{10}{\giga\year}$ and no recent burst. Bottom left: varying the burst age, assuming solar metallicity, $\tau = \SI{10}{\giga\year}$, $\tau_b = \SI{100}{\mega\year}$ and $f_b = \SI{1}{\percent}$. Bottom right: varying the dust attenuation, assuming solar metallicity, $\tau = \SI{10}{\giga\year}$, and no recent burst.}
\label{paperclip}
\end{figure*}

Using these models, we perform a few basic tests to investigate to which parameters our narrowband filters are most sensitive. We calculate the \ac{ew} in our \hb\ and \oii\ filters, $W_{\mathrm{H}\beta}$ and $W_{3746}$, respectively, through all of the models.\footnote{We note again for clarity that $W_{3746}$ does not refer to the equivalent width of an individual absorption line; $3746$ refers only to the central wavelength of the filter.} Then holding all parameters constant, we vary each parameter and see how the \acp{ew} vary as a function of that parameter in our diagnostic $W_{3746}$ vs.\ $W_{\mathrm{H}\beta}$ plots. Figure~\ref{paperclip} shows these plots for four different varying parameters: the $e$-folding timescale of the main population, $\tau$, the metallicity, $Z$, the age of a recent burst, $t_b$, and the $V$-band dust attenuation in the \ac{ism}, $A_{V,\text{ISM}}$. 

Broadly speaking, in each plot as a function of age, $W_{\mathrm{H}\beta}$ and $W_{3746}$ numerically decrease (stronger absorption in \hb; less flux in the \oii\ on-band than the off-band) at very young ages, $\SI{10}{\mega\year} \lesssim t \lesssim \SI{600}{\mega\year}$. When the models reach \SI{600}{\mega\year} ages, both \ac{ew} measurements begin to numerically increase again, with $W_{\mathrm{H}\beta}$ increasing at a faster rate than $W_{3746}$. The age of \SI{600}{\mega\year} is approximately the lifetime of an A main-sequence star, for which the Balmer absorption lines are strongest. Since our \oii\ filters are heavily sensitive to the higher-order Balmer lines, this produces a similar effect in $W_{3746}$. 

The top left plot of Figure~\ref{paperclip} shows how these quantities vary as we allow the $e$-folding timescale of the main population to change. We see that regardless of the timescale, a young population will follow the same track in the diagnostic plot until reaching an age of \SI{600}{\mega\year}. Shorter star-formation timescales (i.e., short main bursts) that are not continually replenishing their supply of A-type stars rapidly have their $W_{\mathrm{H}\beta}$ diminish. Meanwhile, as the higher-order Balmer lines give way to weak absorption lines in our \oii-on band, molecular bands (such as \ch{CN}) contributed by the dominating presence of cooler stars create large absorption features in our \oii-off band, having the effect of creating artificial ``emission'' in $W_{3746}$ (i.e., $W_{3746} > 0$). In contrast, larger $e$-folding timescales (more constant star formation) have more young stars at late times, which result in more absorption in the Balmer lines, resulting in more negative \acp{ew} in both filters even at late times. 

Adjusting the metallicity of the stellar populations (top right of Figure~\ref{paperclip}), we see that the strengths of $W_{\mathrm{H}\beta}$ and $W_{3746}$ change in well-understood ways \citep[e.g.][]{gonzalezdelgado1999,gonzalezdelgado2005}. At ages younger than \SI{1}{\giga\year}, the strengths of the Balmer lines has only a small dependence on metallicity caused by the dependence of stellar evolution in the integrated light of a stellar population. Again, the stellar metallicity of M101 is slightly subsolar \citep{lin2013}, but the subsolar and solar metallicity tracks greatly differ only for very young ages, $\lesssim \SI{100}{\mega\year}$. At ages older than \SI{1}{\giga\year}, the Balmer line metallicity-dependence is stronger. Again, in addition to the higher-order Balmer lines, the $W_{3746}$ measure also feels the increasing effect of molecular bands at old ages which are slightly elevated at higher metallicities. 

Allowing for a recent burst of star formation (bottom left of Figure~\ref{paperclip}) creates strong variations in the strengths of $W_{3746}$ and $W_{\mathrm{H}\beta}$. In these models, the main population is always older than the burst population. It is clear that very recent bursts dominate the light from the old population and approach the characteristics of a population dominated by younger stars. This is primarily attributed to the addition of more A-type stars, deepening the Balmer lines, leading to stronger signatures in $W_{3746}$ and $W_{\mathrm{H}\beta}$. 

Finally, including the effects of dust attenuation in the \ac{ism} can affect the \acp{ew}, particularly at young ages when stars are presumed to be in their birth clouds (bottom right of Figure~\ref{paperclip}). The effect on $W_{\mathrm{H}\beta}$ is strongest for young ages, where dust greatly weakens the strength of the absorption line. In the models of \citet{charlot2000}, young stars are embedded in heavily-extinguished birth clouds in addition to the dust in the \ac{ism} leading to the absorption line weakening we see for young ages. Since all Balmer lines are plagued by this effect to some degree, we see a similar but smaller trend in $W_{3746}$. At old ages, the Balmer lines are slowly becoming weaker while age-independent molecular bands start to contribute, keeping the $W_{3746}$ relatively constant. Both of these trends are explored in more detail by \citet{macarthur2005} in the context of the Lick indices and $D_{4000}$ measurements. 

Briefly summarizing the models, we see that we are very sensitive to some physical properties while being relatively insensitive to others. Among those that we are most sensitive to are differences between $e$-folding timescales of the main population and recent bursts of star formation. As we will discuss in the next section, recent bursts may explain stellar populations that lie in between the two ``prongs'' predicted by the main population. We are less sensitive to the effects of metallicity and dust. Both are constrained by previous measurements \citep{boissier2004,lin2013}, although dust may have a similar scattering effect to a recent burst of star formation where any local dustier regions might artificially weaken the absorption lines we measure. Again, however, since we have masked the youngest, dustiest regions of M101---the star-forming \hii\ regions---these effects should be relatively small.

\section{The Stellar Ages of the Main Disk}\label{sec:main_disk}

In this section, we begin by comparing the stellar population models generated with \ac{cigale} to the binned data inside the main disk (including what we have defined as the inner disk) to investigate any broad age trends present. Then, we examine how the stellar population ages differ between different environments, namely between the spiral arms and interarm regions. As a reminder, we define the main disk of M101 to be those regions inside \ang{;;430} (\SI{14.4}{\kilo\parsec}) of the center (\S\ref{sec:masks}). 

\begin{figure*}
\plotone{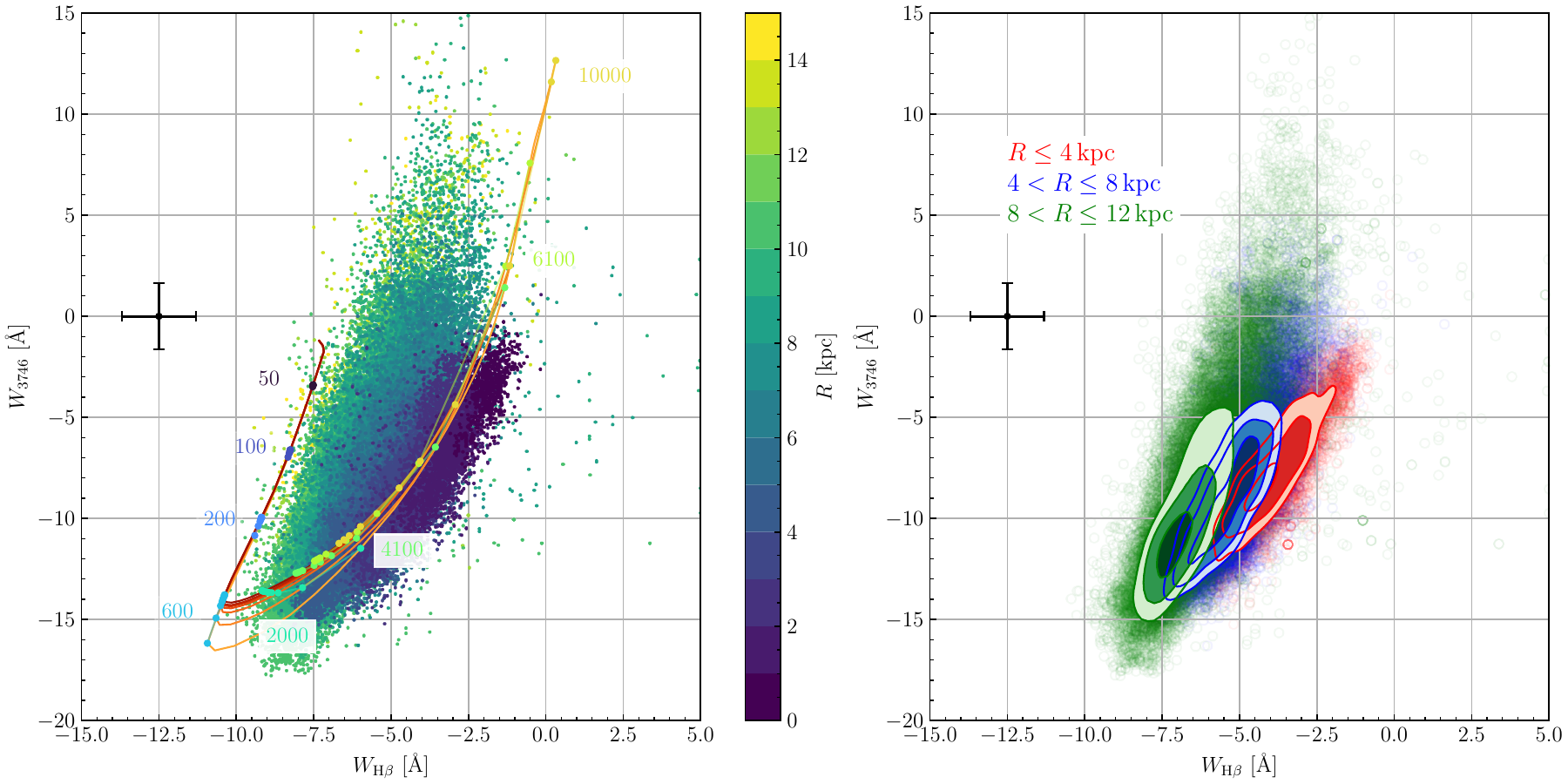}
\caption{$W_{3746}$ vs.\ $W_{\mathrm{H}\beta}$ for all medianed pixels in the main disk of M101 ($R < \ang{;;430}$, $\sim$\SI{15}{\kilo\parsec}). Left: Points are colored by distance from the center of M101. Standard uncertainty shown in the upper left. Overplotted are different models from \ac{cigale}. Models assume a \citet{chabrier2003} \ac{imf}, solar metallicity, and $A_{V,\text{ISM}} = 0.3$. Different colored lines indicate different $\tau$ models: \SIlist{0.5;1;2;3;4;5;10}{\giga\year}. Ages of the stellar populations are indicated on the plot. Right: Contours are colored by distance from the center of M101: $R \leq \SI{4}{\kilo\parsec}$ in red, $\SI{4}{\kilo\parsec} < R \leq \SI{8}{\kilo\parsec}$ in green, and $\SI{8}{\kilo\parsec} < R \leq \SI{12}{\kilo\parsec}$ in blue. Characteristic error bars are shown in the top left. Contours show the data point density for different radial ranges, encompassing \SI{30}{\percent}, \SI{50}{\percent}, and \SI{80}{\percent} of all data points in each range.}
\label{inner_all}
\end{figure*}

Figure~\ref{inner_all} shows our diagnostic $W_{3746}$ v.\ $W_{\mathrm{H}\beta}$ plot for those binned pixels in the main disk of M101 colored by radial distance from the center. Also shown is the prediction from \ac{cigale} models where different star formation histories should fall on this plot. Here, the colored lines show different \ac{sfh} models assuming different $e$-folding times of the main stellar population, ranging from \SI{500}{\mega\year} in light orange to \SI{10}{\giga\year} in red. Points along the tracks are labeled by the age of each model along the track. These models also include the effects of dust assuming an average value of $A_{V,\text{ISM}} = 0.3$ \citep{lin2013}. 

Comparing the data with the generated models generally shows that the inner regions within a few \si{\kilo\parsec} are consistent with an older stellar population, while the mean age generally becomes younger at larger radii. This is more easily seen in the right plot of Figure~\ref{inner_all}, where we plot three radial ranges with density contours. Thus we see that those points at large radii are truly younger with very little old populations mixed in, and vice versa for the points at small radii. Thus, we recover a radial age gradient for the main disk of M101. 

We also see that the two ``prongs'' of the models can be loosely considered the bounds of ``reasonable'' stellar populations. However, a small number of points scatter well outside the model tracks: one plume at high $W_{3746}$ and $W_{\mathrm{H}\beta} \sim \SI{-5}{\angstrom}$ and another at high $W_{\mathrm{H}\beta}$ and $W_{3746} \sim \SI{-5}{\angstrom}$, and a large population of points between the two ``prongs.'' Investigating the locations of the two plumes on our images shows that these are caused by either an unmasked foreground star or scattered diffuse light from masked \hii\ regions.

Most of the true scatter lies between the two ``prongs'' of the model. Returning to Figure~\ref{paperclip}, we see the effects of a recent burst is to bend the older ``prong'' inwards towards the younger ``prong.'' As mentioned before this can be physically attributed to the addition of A-type stars, deepening the Balmer absorption lines. The addition of O/B stars will also weaken the slope of the blue continuum, increasing the value of $W_{3746}$ despite there also being a host of A-type stars as well. Therefore, recent bursts will move stellar populations slightly away from the bend in the ``prongs'' in our models. 

There are also a large amount of points ``in emission'' in $W_{3746}$ in Figure~\ref{inner_all}. As mentioned previously (\S\ref{sub:pop_explain}), this is not caused by any real emission feature being measured, but rather flattening continuum levels in our \oii\ filters combined with molecular bands such as \ch{CN} appearing in our \oii\ off-band filter, creating a false ``emission'' measurement. Recent bursts will bend the older ``prong'' inwards towards being ``in absorption'' in $W_{3746}$ caused by the presence of A-type stars. Given the globally constant and locally stochastic star-forming nature of spiral galaxies, we then expect that pixels dominated by old populations but with a somewhat recent burst (i.e., $t_b \sim \SI{500}{\mega\year}$) will populate this area characterized by being ``in emission'' in $W_{3746}$.

\subsection{Inner Disk vs.\ Arm vs.\ Interarm} 

\begin{figure}
\plotone{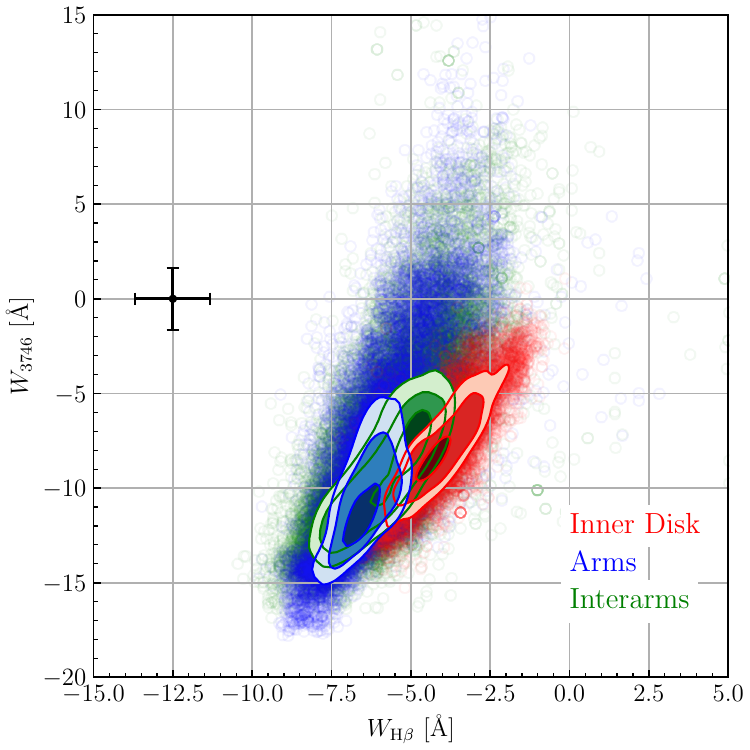}
\caption{The $W_{3746}$ vs.\ $W_{\mathrm{H}\beta}$ for all pixels in the main disk of M101 ($R < \ang{;;430}$, $\sim$\SI{15}{\kilo\parsec}) distinguished by environment: the inner disk in red, spiral arms in blue, and interarm regions in green. Characteristic error bars are shown in the bottom right. Contours show the data point density for different environments, encompassing \SIlist{30;50;80}{\percent} of all data points in each category.}
\label{inner_env}
\end{figure}

We now turn our attention to how the different main disk environments (inner disk, spiral arms, interarm) are spectrally distinct. Applying the masks described in Section~\ref{sec:masks}, Figure~\ref{inner_env} shows how these environments distribute themselves on our $W_{3746}$ v.\ $W_{\mathrm{H}\beta}$ plot. Here, each environment is shown in a different color with density contours overlayed. The inner disk has a tendency to separate itself from the more actively star-forming spiral arms and interarm regions. Notably, there is also a distinct offset between the centroids of the density distributions for the spiral arms and interarm regions. 

These differences, just as in Figure~\ref{inner_all}, can be broadly interpreted as age differences: the inner disk has a stellar population that is older than that in more actively star-forming arm and interarm regions. However, it is worth mentioning that what we define to be the inner disk is much larger than is traditional since we defined the inner disk based on the inability to discern separate spiral arms. This might cause smaller changes in the stellar ages to be unresolvable. For instance, \citet{lin2013} found that the bulge of M101 ($R \lesssim \ang{;;20}$) has a younger stellar age than the surrounding disk by $\sim$\SI{3}{\giga\year} using pixel-based multiwavelength \ac{sed} fitting. Young bulges are found in numerous late-type galaxies \citep[e.g.][]{ganda2007,peletier2007}, and M101 does have a mildly star-forming pseudobulge \citep{fisher2009,fisher2010,kormendy2010}. However, given the extensive network of \hii\ regions in the inner \si{\kilo\parsec} of M101, this region is largely masked in our imaging and thus not measured by our analysis.

Moving on to the spiral arms and interarm regions, the spiral arms are dominated by a recently-formed young population while the interarm region is slightly mixed in ages. The interarm has a sizable young population that overlaps with the spiral arm distribution, but also stellar populations that have an older mean age, more similar to the non-star-forming populations of the inner disk. The observation that the interarm region has a mix of different stellar populations prompts the question of if these populations are spatially distinct. 

\begin{figure}
\plotone{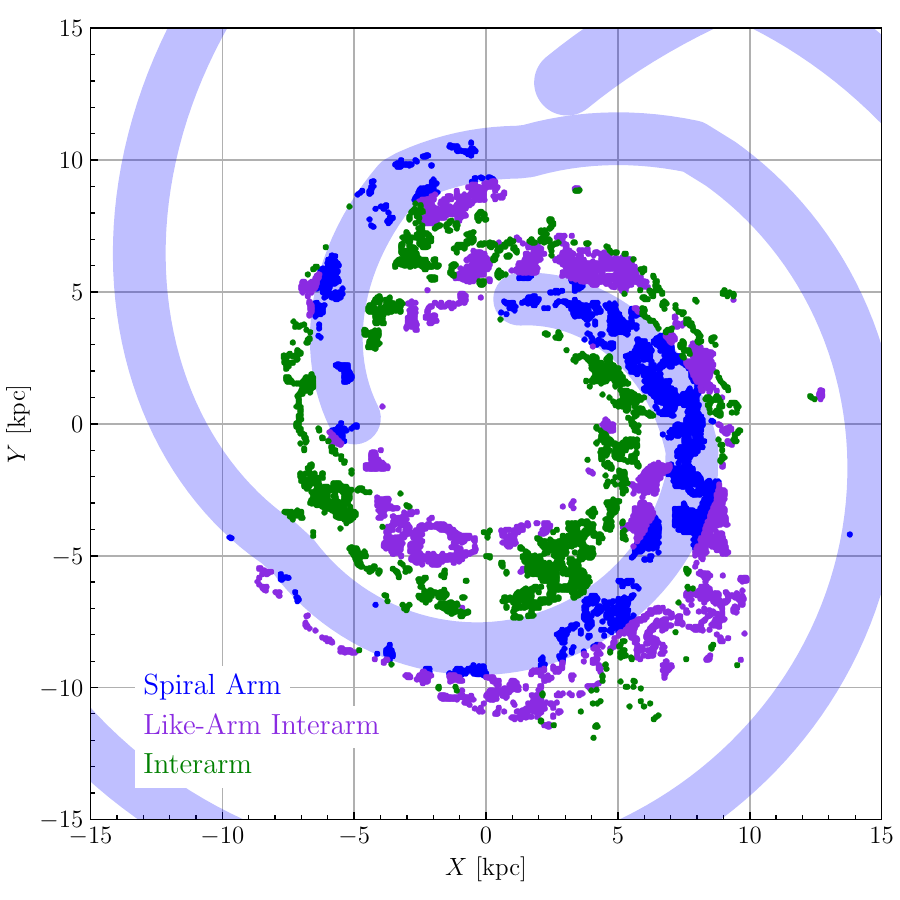}
\caption{The spatial distribution of spectrally-selected regions (see text). The points labeled ``interarm'' are those that are spectrally distinct from the spiral arms, while those labeled ``like-arm interarm'' are those that are spectrally indistinct from the spiral arms. Schematics of the spiral arms are the light blue regions.}
\label{inner_space}
\end{figure}

To test this, we select two spectral regions on our diagnostic plot. We select interarm points that are spectrally distinct from the majority of the arm points, that is $\SI{-5}{\angstrom} \leq W_{\mathrm{H}\beta} \leq \SI{-4}{\angstrom}$ and $\SI{-7}{\angstrom} \leq W_{3746} \leq \SI{-5}{\angstrom}$, and arm and interarm points that are spectrally indistinguishable, that is $\SI{-7}{\angstrom} \leq W_{\mathrm{H}\beta} \leq \SI{-6}{\angstrom}$ and $\SI{-13}{\angstrom} \leq W_{3746} \leq \SI{-10}{\angstrom}$. These points are located approximately where the densest contours lie in Figure~\ref{inner_env}. Thus, the interarm region is split into ``like-arm'' interarm regions and ``true'' interarm regions, defined by their spectral properties. Figure~\ref{inner_space} shows the spatial distribution of these spectrally-defined regions. 

What is immediately apparent is that the ``like-arm'' interarm regions are preferentially found on the outer edge of the spiral arm, while the ``true'' interarm regions are found in the inner edge of the spiral arm.  According to our stellar population models, the spiral arm and ``like-arm'' interarm regions have similar stellar ages, while the ``true'' interarm regions are older. Thus, we observe a potential age gradient in the ages of stars across a spiral arm. There are small areas of ``like-arm'' interarm regions that appear to be continuations of a spiral arm, particular towards the inner disk. This feature is likely caused by the difficulty of defining spiral arms in such a tightly-wound environment. 

\begin{figure*}
\plotone{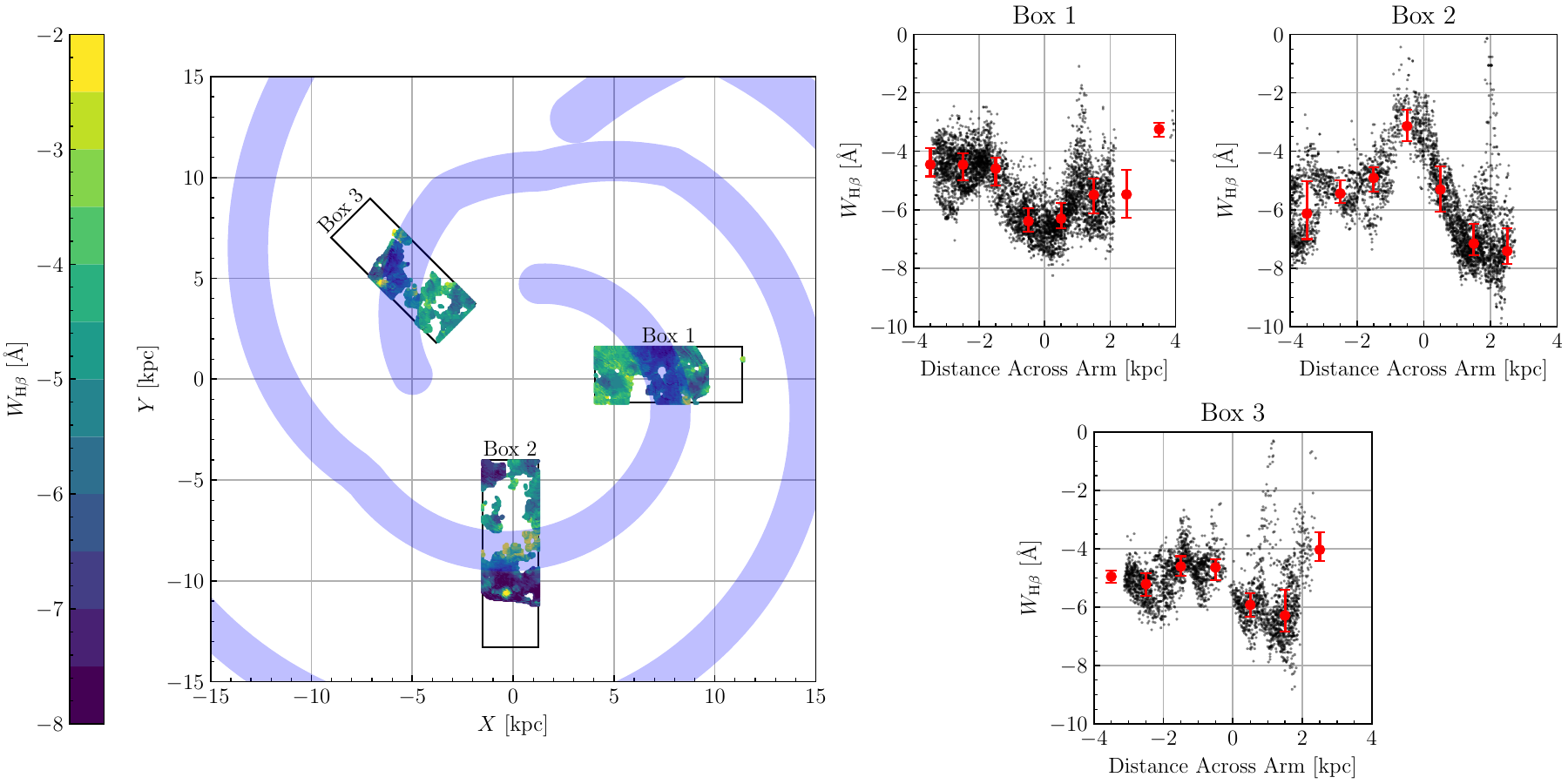}
\caption{The spectral characteristics of spatially-selected regions. Left: three selected boxes in the main disk. Points are colored by their $W_{\mathrm{H}\beta}$. Empty regions in the boxes are either caused by masked \hii\ regions or are below our S/N cut. Right: each box's $W_{\mathrm{H}\beta}$ as a function of distance across a spiral arm. Negative distances are towards the center of M101 and positive distances are away from the center of M101. Binned medians are shown in red. In all three boxes, the $W_{\mathrm{H}\beta}$ is stronger on the outer edge of an arm ($1 \lesssim X \lesssim 3$) than the inside of the arm ($-3 \lesssim X \lesssim -1$), indicative of the shift towards younger populations ahead (moving out of) the arm.}
\label{across_spiral}
\end{figure*}

While Figure~\ref{inner_space} shows the spatial distribution of points selected by their spectral characteristics, an alternative analysis is to look at the spectral characteristics of points along a track cutting through the spiral arms. To check this, in Figure~\ref{across_spiral} we select three rectangular regions extending from outside the inner disk environment, extending through one spiral arm, and ending about halfway to the next spiral arm (i.e., in the middle of an interarm environment). Investigating the trend of $W_{\mathrm{H}\beta}$ with distance across a spiral arm, we see the expected age characteristics: inside the arm lies weak absorption produced by older stellar populations, while moving outside the arm is associated with the strengthening absorption of younger stellar populations. Clearly these conclusions are not only supported by the \ac{sfh} modeling as in Figure~\ref{inner_space}, but also supported observationally with direct measurements of the age-sensitive $W_{\mathrm{H}\beta}$ as in Figure~\ref{across_spiral}. 

Interpreting these observed age trends in the context of spiral density waves is relatively straightforward due to the ordered nature of the main disk of M101. These age trend are well described by the large-scale shock scenario \citep{roberts1969,dixon1971}, which predicts that inside of corotation, ages should change along these cuts through the arm, in this case decreasing along our chosen direction. Thus our observations give strong support for the dynamical scenario where the main disk is characterized by a quasi-steady global spiral wave. As mentioned in the Introduction, this age trend is often hard to measure due to the use of broadband colors and the well-known age-metallicity degeneracy. However, techniques that control for the reddening effects of dust and metallicity, such as the optical/infrared photometric index used by \citet{gonzalez1996} and \citet{martinezgarcia2009}, have started to reveal these age trends across spiral arms. 

Similarly in our data, our ability to detect these age differences may be due to the reduced sensitivity to dust and metallicity in our narrowband imaging compared to broadband optical studies. As mentioned earlier (\S\ref{sub:pop_explain}), $W_{3746}$ and $W_{\mathrm{H}\beta}$ have different responses to dust and metallicity: $W_{\mathrm{H}\beta}$ is only sensitive to one Balmer line that strongly weakens for older populations, while $W_{3746}$ is sensitive to many Balmer lines and the shape of the blue continuum which strongly weakens for younger populations. The combination of these two age indicators allows us to break the degeneracies any one indicator might have \citep{macarthur2005}.

\section{The Stellar Ages of the Outer Disk}\label{sec:outer_disk}

Studying the stellar component of galaxy outskirts is particularly challenging. The surface brightness of the outer stellar component is usually well below that of the night sky, requiring very deep and accurate photometry to constrain the properties of the stellar populations \citep[e.g.][]{mihos2013,watkins2016,peters2017}. In addition, the \acl{lsb} outer regions can be contaminated by background sources and instrumental scattered light, complicating the measurement of photometric \acp{ew}. To alleviate these issues, in our analysis of the outer disk, we focus on a small sample of relatively ``clean'' regions in the outer disk, and compare their properties and distribution to the main disk. These regions were picked by eye, marking areas that are free of contamination and have high enough surface brightness to be well-detected in our narrowband imaging. Each region measures $40 \times 40$ pixels ($\ang{;;58} \times \ang{;;58}$ or $\SI{2}{\kilo\parsec} \times \SI{2}{\kilo\parsec}$), in which we measure the median $W_{3746}$ and $W_{\mathrm{H}\beta}$. 

\begin{figure*}
\plotone{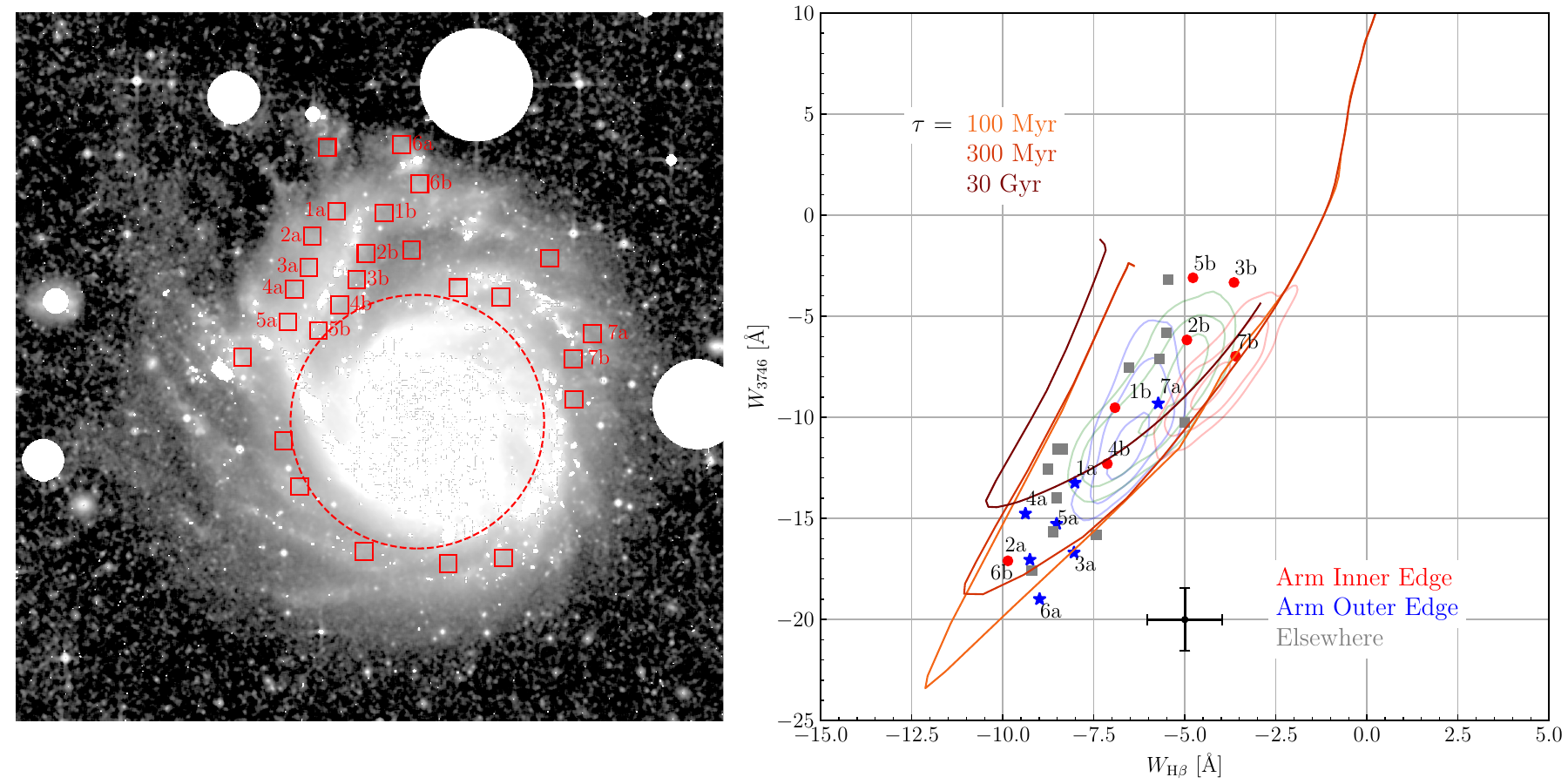}
\caption{Left: The median binned \hb\ off-band image of M101. \hii\ regions and foreground stars are masked. The red dashed circle marks the main/outer disk distinction, $R = \ang{;;430}$. Selected regions in the outer disk are marked by red boxes measuring $40 \times 40$ pixels ($\ang{;;58} \times \ang{;;58}$ or $\SI{2}{\kilo\parsec} \times \SI{2}{\kilo\parsec}$) and numbered. The image measures $\ang{;40;} \times \ang{;40;}$ or $\SI{80}{\kilo\parsec} \times \SI{80}{\kilo\parsec}$. North is up, east to the left. Right: The selected regions' distribution on the diagnostic plot. Labels are carried over from the left panel. Colored points and text (lower right) correspond to the different environments of the regions selected. The colored contours correspond to the density distribution of different environments in the main disk; same as Figure~\ref{inner_env}. Colored tracks and text (upper left) correspond to different \acp{sfh} with different $\tau$. Characteristic error bars are shown in black at the bottom right.}
\label{outer_pos}
\end{figure*}

Figure~\ref{outer_pos} shows the location of each region in the $9 \times 9$ median \hb\ off-band image. Particular pairs of regions inside (`a') and outside (`b') spiral arms are labeled. Regions 1-5 lie along the northeastern arm, chosen to sample stellar populations in the direction of the NE Plume \citep{mihos2013}. Regions 6a and 6b sample either side of a star-forming region, and regions 7a and 7b sample the edge of a short spiral arm to the northwest. Other regions have been placed targeting interarm regions, star-forming complexes, and the faint, outer reaches of the galaxy. In this way, we have attempted to target different features in M101's outer disk, while also remaining azimuthally unbiased.

Figure~\ref{outer_pos} also shows these regions' locations on the diagnostic $W_{3746}$ vs.\ $W_{\mathrm{H}\beta}$ plot. Comparing these outer disk regions with the main disk environments, we see that while there is overlap, many regions show stronger absorption features in one or both of our filters than the main disk. Notably many of these regions are those that are along the northeast arm. Stronger absorption is indicative of a more dominant young population than that in the main disk, suggesting recent star formation in these regions perhaps as a result of the M101-NGC~5474 interaction. Similar studies using broadband colors have also inferred younger ages in the outer disk \citep{bianchi2005,lin2013,mihos2013}. 

Also shown in Figure~\ref{outer_pos} are three representative \ac{sfh} tracks generated with \ac{cigale}. All assume a \citet{chabrier2003} \ac{imf}, solar metallicity, and $A_{V,\text{ISM}} = 0.5$ and differ in the $e$-folding timescale, $\tau$, being \SI{100}{\mega\year}, \SI{300}{\mega\year}, or \SI{30}{\giga\year}. These are single-exponential \acp{sfh} meant to describe the bulk of the stellar population, i.e., the $\tau = \SI{100}{\mega\year}$ model approximates a stellar population that had most of its star formation early and is fading out. The effect of a shorter $\tau$, i.e., a ``burstier'' \ac{sfh}, is to produce stronger absorption lines at the turn-around age of $\sim$\SI{600}{\mega\year}. The different tracks rejoin one another at late times (several \si{\giga\year}).

Comparing the selected outer disk regions with these \ac{sfh} tracks, we see that none of them appear consistent with the ``burstiest'' model, that with $\tau = \SI{100}{\mega\year}$, suggesting a slightly more extended \ac{sfh}. Meanwhile many of the regions on the north and northeastern side of M101 are consistent with young populations. For instance, several regions could be young populations associated with a fading burst (among them regions 2a, 3a, 6a, 6b). Others might be young populations in a near-constant \ac{sfh} or a fading burst (among them regions 4a, 5a). 

Interesting to note are the regions in the direction of the NE Plume beyond the northeast arm (regions `a' 1-5). These five regions form a tight cluster of points in Figure~\ref{outer_pos}, near the turn-around point in the \ac{sfh} tracks. The fact that they have strong absorption features through our filters suggests that these are young ages, consistent with a recent (few hundred \si{\mega\year}) burst of star formation on top of a somewhat older but not dominant background stellar population. This timescale is similar to that found by \citet{mihos2018} who used \acs{hst} to study the discrete stellar populations in M101's NE Plume (at a slightly larger radius than probed here), finding evidence for both an old population and a $\sim$\SI{300}{\mega\year} starburst population in the region. In our studies of the integrated light here, the starburst population and older background populations combine to give an older mean age (\SI{500}{\mega\year}-\SI{1}{\giga\year}), but the scattering of regions shows good agreement with the picture of a recent weak starburst in this part of M101's outer disk. 

However, Figure~\ref{outer_pos} also shows other regions in M101's outer disk which are dominated by older populations. Some of these lie in interarm regions or to the west of M101, a part that likely did not participate strongly in the M101-NGC~5474 interaction (in the model of \citealt{linden2022}, these regions were on the opposite side from the companion at closest approach). Interestingly, most of the selected regions on the inside of spiral arms (labeled `b' in Figure~\ref{outer_pos}) have weaker absorption features, suggesting an older mean age than their partners on the outside of the spiral arms. These age patterns across the outer arms are similar to those seen in the main disk, again indicative of the large-scale shock scenario. However, here we are sampling regions well outside the corotation radius of the main disk (found to be at \SI[multi-part-units=single]{15.6 \pm 2.2}{\kilo\parsec}; \citealt{roberts1975,scarano2013}). Outside of corotation, we would expect the age gradients to be flipped (older populations leading the arms and younger populations inside; \citealt{roberts1969,dixon1971}). The fact that we observe no change in the directionality of the age gradients in the outer arms hints at the possibility of multiple pattern speeds in M101, which we explore below.

\section{Multiple Pattern Speeds of M101?}

In the past two sections, we have presented evidence of the large-scale shock scenario in both the main and outer disks of M101. There exists age gradients across the spiral arms of M101 going from primarily old populations on the inner edge to younger populations along the outer edge of an arm. This gradient is predicted by the large-scale shock scenario \citep{roberts1969,dixon1971}, but interestingly we see no reversal of the gradient outside M101's main corotation radius ($R \approx \SI{15}{\kilo\parsec}$; \citealt{scarano2013}) contrary to the expectation in the large-scale shock scenario. 

Thus there are two possible responses: either star formation patterns in M101 are not governed by the large-scale shock scenario, or there exists multiple pattern speeds, and thus multiple corotation radii, producing the trends we observe. The large-scale shock scenario requires star formation to occur preferentially in the spiral arms, producing age gradients across the arms that flip direction inside and outside corotation. Conversely, if star formation occurs stochastically throughout the disk regardless of the location of the spiral arms, this could introduce spurious patterns in the age distributions that are inconsistent with the large-scale shock scenario. While some studies have questioned the importance of spiral arms in driving star formation \citep{foyle2010,foyle2011,ragan2018,querejeta2021}, that does not appear to be the case in M101. Not only do we see here the gradients predicted by large-scale shocks, studies of the galaxy's massive star populations show strongly enhanced star formation in the spiral arms (up to 30 times more efficient than in the interarm regions; \citealt{cedres2013}). This picture of spiral-driven star formation is echoed in studies of other spiral galaxies \citep{vogel1988,lord1990,knapen1996}, and thus, at least for M101, the large-scale shock scenario likely still holds. 

Instead, what is more probable is that M101 is a spiral galaxy with multiple pattern speeds. The existence of multiple pattern speeds which dominate over concentric radial ranges has been measured for M101 and other galaxies in the past. \citet{meidt2009} used \hi\ and \ch{CO} data cubes of M101 to measure these pattern speeds and found at least three distinct speeds each with their own corotation radius: an inner pattern within \SI{6}{\kilo\parsec}, a second extending from \SIrange{6}{13}{\kilo\parsec}, and a third from \SIrange{13}{20}{\kilo\parsec}. Each pattern has its own corotation radius: \SIlist[list-final-separator = {, and }]{3.7;9.8;19.6}{\kilo\parsec}, respectively. Using different methods, other studies have confirmed the innermost pattern speed as well \citep{egusa2009,cedres2013}. 

The possible existence of three pattern speeds in M101 could explain the recurrence of the same age trend in the outer disk as in the inner disk. At any particular radii in M101, the gas and stars would always be inside of some corotation radius, resulting in stellar populations that are primarily old along the inner edge of a spiral arm and becoming younger across the arm towards the outer edge. There would likely not be a flipping of this trend until beyond the outermost pattern speed. The fact that we don't see this flip in the outer disk of M101 does suggest that the outermost pattern speed in \citet{meidt2009} does extend further than their cutoff of $\sim$\SI{20}{\kilo\parsec}, potentially to at least \SI{25.5}{\kilo\parsec} (the galactocentric distance to region 1a in Figure~\ref{outer_pos}).

The adoption of multiple pattern speeds necessarily requires the adoption of ``dynamic'' spiral arms (i.e., transient and recurrent spiral arms; e.g.\ \citealt{sellwood1984,elmegreen1986,sellwood2011}) as opposed to quasi-static, long-lived spiral arms \citep[e.g.][]{lin1964,bertin1989}. In the former theory, spiral arms appear and reappear in cycles, breaking up into smaller segments with sizes of a few \si{\kilo\parsec}, then reconnecting by differential rotation to reform large-scale patterns \citep{wada2011}. Thus dynamic spiral arms can appear to be long-lived visually, but change on short timescales. 

A ``dynamic'' M101 naturally has consequences for its shape and spiral structure. Multiple pattern speeds are linked and supported by ``mode coupling'' (\citealt{sygnet1988,masset1997}; see \citealt{sellwood2019} and references therein for more recent discussions), wherein resonances of different pattern speeds overlap, such as the corotation radius of an inner pattern overlapping with the inner Lindblad resonance of an outer pattern. This overlap causes energy and angular momentum to be transferred efficiently to the outer disk, building up the spiral pattern at large radii \citep[e.g.][]{masset1997,rautiainen1999,meidt2008_sim,meidt2008_m51,font2014}. A consequence of mode coupling noted in simulations is that galaxies are stimulated to produce $m=1$ spiral waves \citep[e.g.][]{sellwood1988,rautiainen1999,salo2000}, just as we see with the large asymmetric spiral arm in M101. Notably, this might be a potential solution to the problem \citet{linden2022} faced in constraining the mass of NGC~5474, having to adopt a mass ratio of one-eighth that of M101 leading to a significantly higher circular velocity for the satellite than observed in order to reproduce the asymmetry by interaction alone. If the disk of M101 was already sensitive to producing $m=1$ modes, it could be quite responsive to tidal forcing even with a lower mass for NGC~5474. Fully hydrodynamical simulations of the M101-NGC~5474 interaction, building on the work of \citet{linden2022}, will need to be performed to test this theory. 

However, a ``dynamic'' M101 and our measurements of an age gradient across spiral arms has consequences for computational studies of spiral structure in disk galaxies. In contrast to the canonical density wave model for spiral structure, many hydrodynamical simulations show spiral patterns forming as shearing arms that co-rotate with the stars and gas \citep[e.g.][]{dobbs2010,grand2012,baba2015,baba2017,dobbs2017}. This co-rotation makes it such that molecular clouds, \hii\ regions, and star clusters do not easily move across and out of the spiral arms, predicting that no age gradients should be seen across the arm, in contrast to what we observe in M101. Other models, generate spiral structure via a force response to mass clumps in the disk (\citealt{donghia2013}; see also the introduction of \citealt{sellwood2019}), which produce ``wakes'' that extend over a small radial range and could produce the age gradients we see in M101. However, these wakes do not extend over large radii and thus do not get far from corotation. Such a model would thus predict much milder age gradients across the arms, possibly at odds with the strong age gradients we see in M101. However, we also note that most simulations of dynamic spiral arm mechanisms produce relatively weak spiral structure \citep[e.g.][]{dobbs2010,grand2012,donghia2013}. These models may provide good descriptions of ``flocculent'' spiral structure, but in stark contrast stands M101, a galaxy with strong spiral arms that are dynamic in nature and also likely influenced by M101's recent interaction with NGC~5474. Clearly, new simulations of dynamic spiral structure are needed that reproduce the strong spiral arms of M101 while also resulting in age gradients observed across its spiral arms.

\section{Conclusions}

Using narrowband filters that measure the absorption line strengths of \hb\ and higher order Balmer lines and metallic lines between the Balmer break at \SI{3646}{\angstrom} and the \SI{4000}{\angstrom} break, we have placed constraints on the stellar ages throughout the disk of M101. We calibrate and confirm the efficacy of this technique using the stellar population synthesis code \ac{cigale} \citep{noll2009,boquien2019} to observe the effect of stellar population parameters on equivalent widths measured through our filters. Our narrowband imaging technique proves sensitive to the star formation history of the stellar populations, and is only modestly sensitive to other effects such as dust and metallicity. We divided M101 into several different radial regions to study the mean population age and also examined the differences between spiral arms and their interarm environments. We confirm studies of the overall radial age gradient as well as shown new evidence for spiral driven features in the spatial age distribution. We interpret these results in the context of spiral arm dynamics and multiple pattern speeds in the disk of M101. Our main conclusions are summarized below. 

\begin{enumerate}

\item In the main disk, we focused on the differences between morphological features, i.e., spiral arms, interarm regions, and the inner disk. We confirm previous studies showing a radial age gradient \citep{bianchi2005,lin2013}, where the inner disk has populations older than those at larger radii. Comparing spiral arms and interarm populations revealed a spread in the mean ages of the interarm population, some similar to the young arms, and others older and more akin to the inner disk. These ``like-arm'' interarm regions are preferentially found on the outer edge of a spiral arm, while the ``true'' interarm regions are found on the inner edge. 

\item We interpreted these spatial trends in the context of the ``large-scale shock scenario'' \citep{roberts1969,dixon1971} a consequence of quasi-steady, global spiral density waves. Inside of corotation, gas clouds overtake the spiral arm and form stars in \hii\ regions. These stars drift outwards, creating a color and age gradient across the spiral arms. Despite being historically hard to measure observationally, this evolutionary path has been predicted in simulations \citep[e.g.][]{dobbs2010}, and is now clearly in evidence in our narrowband imaging.

\item To combat the low signal-to-noise in the diffuse outer disk, we selected a number of ``clean'' regions where we can spatially bin over larger areas to accurately measure stellar absorption signatures. Unexpectedly, we recovered a similar age trend across the spiral arms in the outer disk as we had found in the main disk, with younger populations along the outer edge and older populations along the inner edge of spiral arms. The outer disk, while having significantly younger populations than the main disk, did also contain some old populations similar to the main disk. 

\item In terms of processes shaping the outer disk, the lack of a change in the spiral arm age trends beyond the assumed corotation radius of M101 suggests that there are multiple pattern speeds and thus multiple corotation radii in M101. Other groups have found evidence for radially varying pattern speeds in M101 and other galaxies \citep{meidt2009,font2014}, and our work supports this conclusion. 

\end{enumerate}

Overall, our results are consistent with a picture where M101 is a ``dynamic'' galaxy, one with transient and recurrent spiral arms. While still showing the radial signs expected by inside-out galaxy formation, where the inner region is older and the outer region is younger, the nature of the spiral pattern does not conform to standard density wave theory as measured by stellar ages. Instead, there may be multiple pattern speeds in M101 linked by mode coupling. This allows resonances in the different spiral patterns to overlap, causing energy and angular momentum to be transferred efficiently to the outer disk, building up the spiral pattern and resulting in the one-armed spiral pattern we see in M101. While this has promising answers for the trends in M101 seen here and elsewhere, hydrodynamical modeling of M101 and its interaction history needs to be performed to fully unravel these mysteries.

\begin{acknowledgments}

The authors would like to thank Aaron Watkins for his work in collecting the \ha\ image used for determining the widths of the spiral arms. We thank Stacy McGaugh for helpful discussions and comments on this paper. We also thank Charley Knox for his work supporting the Burrell Schmidt, and the Mt.\ Cuba Astronomical Society for financial support for the project. We also thank the anonymous referee for a detailed report that helped improve this paper. R.G.\ was supported in part by a Towson Memorial Scholarship. R.G.\ would like to thank \textsc{StackOverflow} user \texttt{ImportanceOfBeingErnest} for their \texttt{Python} code used in drawing the spiral arm masks. This publication makes use of data products from the Two Micron All Sky Survey, which is a joint project of the University of Massachusetts and the Infrared Processing and Analysis Center/California Institute of Technology, funded by the National Aeronautics and Space Administration and the National Science Foundation. This research has made use of the NASA/IPAC Extragalactic Database, which is funded by the National Aeronautics and Space Administration and operated by the California Institute of Technology. 

\end{acknowledgments}

\emph{Facility:} CWRU:Schmidt

\software{\texttt{Astropy} \citep{astropy2013,astropy2018}, \texttt{Matplotlib} \citep{matplotlib}, \texttt{NumPy} \citep{numpy}, \acs{cigale} \citep{noll2009,boquien2019}, \texttt{SciPy} \citep{Virtanen2020}, \texttt{piecewise-regression} \citep{piecewise}, \texttt{SAOImage ds9} \citep{ds9}, \texttt{Phyla/adjustText} \citep{flyamer2023}}

\bibliography{M101_Ages.bib}
\bibliographystyle{aasjournal.bst}

\end{document}